\documentclass[]{mn2e}

\usepackage{amssymb}
\usepackage{amsmath}
\usepackage{graphicx}
\usepackage{color}
\usepackage{longtable}
\usepackage{url}

\title[$\phi(M_*)$ and variable-IMF]{Stellar mass functions and implications for a variable IMF}
\author[Bernardi et al.]{\parbox{\textwidth}{M. Bernardi$^{1}$\thanks{E-mail: bernardm@sas.upenn.edu}, R. K. Sheth$^{1}$, J.-L. Fischer$^{1}$, A. Meert$^{1}$, K.-H. Chae$^{1,2}$, H. Dominguez-Sanchez$^{1}$, M. Huertas-Company$^{1,3}$, F. Shankar$^{4}$ \& V. Vikram$^{1}$} \vspace{0.4cm}\\
\parbox{\textwidth}{$^{1}$Department of Physics and Astronomy, University of Pennsylvania, Philadelphia, PA 19104, USA\\
$^{2}$ Department of Astronomy and Space Science, Sejong University, 98 Gunja-dong Gwangjin-Gu, Seoul 143-747, Republic of Korea\\ 
$^{3}$GEPI, Observatoire de Paris, CNRS, Univ. Paris Diderot;
Place Jules Janssen, 92190 Meudon, France\\
$^{4}$Department of Physics and Astronomy, University of Southampton,
Southampton SO17 1BJ, UK\\}}

\begin{document}
 \date{Accepted .  Received ; in original form }

\maketitle

\label{firstpage}

\begin{abstract}
  Spatially resolved kinematics of nearby galaxies has shown that the ratio of dynamical- to stellar population-based estimates of the mass of a galaxy ($M_{*}^{\rm JAM}/M_{*}$) correlates with $\sigma_e$, the light-weighted velocity dispersion within its half-light radius, if $M_{*}$ is estimated using the same Initial Mass Function (IMF) for all galaxies and the stellar mass-to-light ratio within each galaxy is constant.  This correlation may indicate that, in fact, the IMF is more bottom-heavy or dwarf-rich for galaxies with large $\sigma$.  We use this correlation to estimate a dynamical or IMF-corrected stellar mass, $M_{*}^{\rm \alpha_{JAM}}$, from $M_{*}$ and $\sigma_e$ for a sample of $6 \times 10^5$ SDSS galaxies for which spatially resolved kinematics is not available.  We also compute the `virial' mass estimate $ k(n,R)\,R_e\,\sigma_R^2/G$, where $n$ is the S{\'e}rsic index, in the SDSS and ATLAS$^{\rm 3D}$ samples.  We show that an $n$-dependent correction must be applied to the $k(n,R)$ values provided by Prugniel \& Simien (1997). Our analysis also shows that the shape of the velocity dispersion profile in the ATLAS$^{\rm 3D}$ sample varies weakly with $n$: $(\sigma_R/\sigma_e) = (R/R_e)^{-\gamma(n)}$.
  The resulting stellar mass functions, based on $M_*^{\rm \alpha_{JAM}}$ and the recalibrated virial mass, are in good agreement.  If the $M_*^{\rm \alpha_{JAM}}/M_{*} - \sigma_e$ correlation is indeed due to the IMF, and stellar mass-to-light gradients can be ignored, then our $\phi(M_{*}^{\rm \alpha_{JAM}})$ is an estimate of the stellar mass function in which $\sigma_e$-dependent variations in the IMF across the population have been accounted for.
  Using a Fundamental Plane based observational proxy for $\sigma_e$ produces comparable results. The use of direct measurements for estimating the IMF-dependent stellar mass is prohibitively expensive for a large sample of galaxies.  By demonstrating that cheaper proxies are sufficiently accurate, our analysis should enable a more reliable census of the mass in stars, especially at high redshift, at a fraction of the cost. Our results are provided in tabular form.

\end{abstract}

\begin{keywords}
 galaxies: luminosity function, mass function -- galaxies: structure -- galaxies: fundamental parameters -- galaxies: kinematics and dynamics

\end{keywords}

\section{Introduction}
The comoving number density of galaxies evolves, and this encodes information about how galaxies formed.  Early work studied the galaxy luminosity function -- the comoving density in bins of luminosity -- but because the luminosity depends on waveband, and can evolve even if the mass in stars does not, the last decade has seen interest shift from the luminosity function $\phi(L)$ to the stellar mass function $\phi(M_*)$ (Bernardi et al. 2017a and references therein).

Stellar masses are typically estimated as the product of the luminosity $L$ and a stellar mass-to-light ratio $M_*/L$. While there are systematics associated with both, a series of recent papers make the case that systematics associated with $L$ have been significantly reduced in recent years (Bernardi et al. 2013; Meert et al. 2015; Fischer et al. 2017; Bernardi et al. 2017a,b) and are now smaller than those for $M_*/L$ \cite{b17a}. 

Current algorithms for determining $M_*/L$ require one to make a number of assumptions about the age, metallicity, star-formation history, dust content, and so on.  However, one of the largest systematics in the determination of $M_*/L$ is the shape of the initial stellar mass function (hereafter IMF); the two most popular choices \cite{salpeter,chabrier} differ by $\sim 0.25$~dex, with Salpeter being heavier, i.e., having larger $M_*/L$, than Chabrier.  The difference between these two is primarily because the Salpeter IMF includes many more low mass stars, as a result of which it is often said to be `bottom-heavy' or `dwarf-rich'.

Moreover, essentially all estimates of $\phi(M_*)$ assume that the IMF is constant across the population.  This assumption is one of convenience -- it has no physical motivation.  Indeed, a number of recent observations -- some based on gravity-sensitive features in the spectrum (Conroy \& van Dokkum 2012; La Barbera et al. 2013; Spiniello et al. 2014; Lyubenova et al. 2016; Lagattuta et al. 2017), others based on gravitational lensing and stellar dynamics (Auger et al. 2010; Thomas et al. 2011; Spiniello et al. 2012; Cappellari et al. 2012; Cappellari et al. 2013b; Barnab\`e et al. 2013; Posacki et al. 2015) -- suggest that the IMF is not constant across the population.  Even more recent work suggests that the IMF is not even constant within a galaxy \cite{califaIMF,cvdNew,slopeIMF}.  The primary goal of the present study is to incorporate the first of these effects into an estimate of $\phi(M_*)$ in a sample, the SDSS DR7, that is two orders of magnitude larger than those in which IMF-variations have been detected directly.  We leave accounting for IMF gradients within a galaxy for future work.

A particularly simple estimate of $\phi(M_*)$, in which the IMF was assumed to depend on morphological type --  Salpeter for early types, and more Chabrier-like for later types -- was made by Bernardi et al. (2010).  Their Figure~25 shows that, in contrast to when the IMF is assumed to be the same for all galaxies, the resulting estimate of $\phi(M_*)$ is rather similar to $\phi(M_{\rm dyn})$, where $M_{\rm dyn} = 5 R_{e,{\rm deV}} \sigma^2/G$ and $R_{e,{\rm deV}}$ is the half-light radius from a de~Vaucouleurs profile. They suggested that IMF-variations may be a reasonable way of reconciling stellar population and dynamical estimates of the stellar mass function.  Subsequent work suggests that the IMF may vary with metallicity \cite{csf16}.  And even more recent work suggests that the IMF varies across the population even when the morphological type is fixed \cite{mangaIMF}.  Moreover, the size estimate $R_e$ depends on the model fitted to the light profile, so there is no compelling reason to use $5 R_e\sigma^2/G$ whatever the fitted model \cite{ps97}.  Therefore, our goal is to present and compare better motivated estimates of $\phi(M_*)$ and $\phi(M_{\rm dyn})$.  

Achieving our goal is complicated by the fact that we do not have spectra with sufficient signal to noise or wavelength coverage to see the IMF-related spectroscopic features directly, so we must use other proxies.  Velocity dispersion is potentially a good choice, because Conroy \& van Dokkum (2012), using IMF-sensitive features in the spectra of 38 early-type galaxies, have shown that $M_{*,\rm IMF}/M_{*}$, the ratio of stellar masses estimated allowing the IMF to vary to that where it is held fixed, correlates with velocity dispersion:  the IMF tends to be bottom-heavy (i.e. dwarf-rich) in galaxies with large velocity dispersions.  Other groups have come to similar conclusions \cite{spiderIMF,califaIMF,fireIMF}, although the agreement is not universal \cite{sl13,salcc15,csf15}.

More recently, using spatially resolved photometry {\em and} spectroscopy of 26 galaxies over a range of scales, Lyubenova et al. (2016) used a Jeans-equation analysis of the data (following methods described in Cappellari et al. 2012) to estimate what we will refer to as $M_{*}^{\rm JAM}$ for each galaxy.  The JAM estimate is a dynamical estimate which explicitly models both stellar and dark matter components; $M_{*}^{\rm JAM}$ is the stellar mass component.  Lyubenova et al. found that, for the 26 galaxies they studied, the ratio $M_{*}^{\rm JAM}/M_{*}$ is similar to $M_{*,\rm IMF}/M_{*}$.  Therefore, $M_*^{\rm JAM}$ is another potential proxy for $M_{*,\rm IMF}$.  Indeed, in their analysis of $\sim 800$ galaxies from the Mapping Nearby Galaxies at APO survey (MaNGA), Li et al. (2017) estimated $M_{*}^{\rm JAM}/M_{*}$ for their sample and simply assumed that $M_{*}^{\rm JAM}/M_{*} = M_{*,\rm IMF}/M_{*}$.  
Unfortunately, estimating $M_*^{\rm JAM}$ requires spatially resolved kinematics, which we do not have.  Hence, one of our goals is to combine the the methodology used by Prugniel \& Simien (1997) with a calibration to the ATLAS$^{\rm 3D}$ sample (Cappellari et al. 2011) to estimate $M_*^{\rm JAM}$. 

Section~\ref{sample} describes our SDSS sample, and the observables available in it which we can use to construct our proxies for the IMF-corrected stellar mass.  One of these -- the dynamical mass -- is the subject of Section~\ref{sec:Mdyn}. This Section includes an analysis of the ATLAS$^{\rm 3D}$ sample which we use to calibrate some of our mass estimates.  Section~\ref{proxies} defines a number of estimates of the total, dynamical and/or IMF-corrected stellar mass; this also includes the use of a Fundamental Plane-based photometric proxy for the velocity dispersion (Section~\ref{sec:fp}).  Section~\ref{phiMs} compares the associated stellar mass functions.  A final section summarizes and places our results in the wider context of how variations in the IMF impact estimates of the stellar mass function when IMF-gradients are ignored.  Systematic effects associated with our fixed IMF $M_*$ estimates, our sample, and comparison with previous work are described in Appendix~\ref{ml}, and Appendix~\ref{tables} provides some of our results in tabular form.

When necessary, we assume a spatially flat background cosmology with parameters $(\Omega_m,\Omega_\Lambda)=(0.3,0.7)$, and a Hubble constant at the present time of $H_0=70$~km~s$^{-1}$Mpc$^{-1}$, as these are the values adopted in most studies of the stellar mass function which we reference in our work.  As we will be working at low $z$, all our conclusions are robust to small changes in these parameters.

\section{The Sample}\label{sample}
This section describes the observables we use to define a number of estimators of the stellar mass of galaxies in the SDSS DR7 Main Galaxy Sample \cite{sdssDR7}.

\subsection{Photometry and morphological type}
We select the galaxies in the SDSS DR7 with r-band Petrosian magnitude limits $14 \le m_r \le 17.77$ mag (see Meert et al. 2015 for a detailed discussion of the sample selection).  As discussed in a series of papers (Bernardi et al. 2013; Meert et al. 2015; Fischer et al. 2017; Bernardi et al. 2017b and references therein) the SDSS pipeline photometry underestimates the brightnesses of the most luminous galaxies. This is mainly because (i) the SDSS overestimates the sky background and (ii) single or two-component S{\'e}rsic-based models fit the surface brightness profile of galaxies better than the de Vaucouleurs model used by the SDSS pipeline, especially at high luminosities.  Therefore, rather than the SDSS pipeline photometry, we use the {\tt PyMorph} photometry of Meert et al. (2015).  The differences between {\tt PyMorph} and SDSS pipeline photometry are significant for the most massive galaxies.  Bernardi et al. (2017b) show that these differences are not dominated by intracluster light.

For single S{\'e}rsic fits, the relevant {\tt PyMorph} parameters are the S{\'e}rsic index $n$, half-light radius $R_e$ and total luminosity $L$ of each of object.  The estimated total light $L$ results from extrapolating the fitted (S{\'e}rsic) model to infinity.  As a result, the single S{\'e}rsic fits are known to slightly over-estimate the total light; integrating out to only $8R_e$ yields a more reliable luminosity estimate \cite{b17a}. For two-component SerExp fits the returned parameters are $n_{\rm Bulge}, R_{e,\rm Bulge}$ and bulge luminosity $L_{\rm bulge}$ for one component, and $R_e$ and $L$ for the sum of the two components (where the second component is forced to have S{\'e}rsic index $n=1$).  (In this case, the difference between truncating and extrapolating to infinity matters less.  See, e.g., Figures~\ref{F0} and~\ref{F1}.) We will also consider single component fits in which {\tt PyMorph} forced $n=4$; we refer to the associated size and luminosity as $R_{e,{\rm deV}}$ and $L_{\rm deV}$.  In this paper we will always use {\tt PyMorph} truncated luminosities.

Later in this paper, we will consider E+S0s separately from the full population.  For this, we use the Bayesian Automated morphological classifications (hereafter BAC) of Huertas-Company et al. (2011); our results do not depend strongly on this choice.  We use the BAC classifications because they provide a probability $p$(type) for each object.  We can either weight by, or implement hard cuts in, this probability.  See Bernardi et al. (2010, 2013, 2014) for discussion of morphology as a function of galaxy mass. 

\subsection{Fixed-IMF stellar masses}
For each galaxy, stellar masses $M_*$ are derived by multiplying the Meert et al. (2015) truncated $L_{\rm trunc}$ values by $M_*/L$ values taken from Mendel et al. (2014).  These $M_*/L$ were estimated by fitting the spectral energy distribution (SED) of synthetic stellar population models to all five SDSS wavebands for all objects. Briefly, Mendel et al. compared each galaxy’s observed SED to a synthetic stellar population models grid. This grid was constructed using the flexible stellar population synthesis code of Conroy et al. (2009) which allows us to generate synthetic SEDs given a range of galaxy properties. These models span a range of ages, metallicities, star-formation histories, and dust properties observed in nearby galaxies. A Chabrier IMF was assumed. The relevant parameters of the models grid are summarized in Table 2 of Mendel et al (2014). Appendix~\ref{ml} shows the impact on $\phi(M_*)$ of varying a number of assumptions about the stellar population (e.g. dusty, dust-free, etc.) while always assuming the IMF is constant (Chabrier) across the population.  Appendix~\ref{ml} also compares the stellar mass function of this work with recent estimates (i.e. Bernardi et al. 2013, 2017a).

\subsection{Velocity dispersion}
The SDSS pipeline provides estimates of the velocity dispersion $\sigma_{\rm a}$ measured within a circular aperture of radius $\theta_{\rm a}=1.5$ arcsec for most but not all objects.  Strictly speaking, $\sigma_{\rm a}$ is a complicated combination of rotation, dispersion and orientation with respect to the line of sight, so treating it as a pure velocity dispersion is at best correct for objects which are not rotating; this is usually the case for the most massive galaxies (e.g. Cappellari et al. 2013a).  Since the fiber typically covers the central regions of a galaxy, it may be that $\sigma_{\rm a}$ primarily samples the bulge component of two-component galaxies.  Therefore, in what follows, we will be careful to consider E+S0s separately before extending our results to all galaxy types.  It should be borne in mind that our methods below are reasonably well motivated only for E+S0s, even though we go on to apply them to other types.  

The official pipeline, {\tt SpecObjAll velDisp}, does not provide estimates if the signal-to-noise of the spectrum was too low, or if the object was not early-type.  In contrast, {\tt galSpecInfo v\_disp} does not implement these cuts.  Although our primary interest is in E+S0s, we are also interested in the full range of galaxy types, so we use these latter values.  Hyde \& Bernardi (2009) show that, when both estimates are available, they tend to agree well.  In practice, $\sigma_{\rm a}$ values smaller than $\sim 70$~km~s$^{-1}$ are quite unreliable.

Unfortunately, because $\sigma_{\rm a}$ is estimated in an aperture of fixed angular size, it samples a distance dependent physical scale in each galaxy.  There is general agreement that it is much better to work with a physical scale which is a fixed multiple of the half-light radius.  In what follows, we will work with $\sigma_e$ and $\sigma_{e/8}$, which are the light-weighted projected velocity dispersion within the projected half-light radius, $R_e$, and within $R_e/8$, respectively.  Our first task will be to estimate $\sigma_e$ and $\sigma_{e/8}$ from $\sigma_{\rm a}$; later on, we describe how we treat objects for which not even $\sigma_{\rm a}$ is available.

The first question is:  How different are the fiber and half-light scales?  For the E+S0s in our sample, $t_a\equiv \theta_a/\theta_e\approx 0.6$ approximately independent of $\sigma_{\rm a}$.  However, it correlates weakly with $M_*$:
$t_a\approx {\rm min}[0.6,0.6 - 0.33\log_{10}(M_*/10^{11}M_\odot)]$ because the more massive galaxies are seen out to larger distances -- where the fixed angular aperture corresponds to a larger physical size -- but this trend is weakened by the fact that massive galaxies also tend to have larger half-light radii.  The main point is that $\sigma_{\rm a}$ is measured on scales that are, on average, of order $2\times$ smaller than $R_e$.  

Previous work suggests the following empirical relation for the scale dependence of $\sigma$:
\begin{equation}
  {\rm log}_{10}\frac{\sigma_R}{\sigma_e}
  = -0.066\, {\rm log}_{10} \frac{R}{R_e}
    - 0.013\,\left[{\rm log}_{10}\frac{R}{R_e}\right]^2
 \label{jfk}
\end{equation}
\cite{jfk95}.  In some more recent work \cite{mehlert03,sauronIV}, the second order term is dropped.  Thus, $\sigma_e$ is expected to be approximately $1.04\times$ smaller than $\sigma_{\rm a}$; a difference of about 0.019~dex.  Of course, we can use this same scaling law to estimate $\sigma_{e/8} \approx 1.12\,\sigma_e$.  Section~\ref{sigcorr} presents a new empirical scaling relation which depends on the shape of the light profile (see equation~\ref{gamman}). 

Section~\ref{sec:fp} addresses the question of how to proceed if $\sigma_{\rm a}$ is not available or is too expensive to measure (e.g. in more distant samples).  It argues that one should be able to build a cheaper observational proxy for $\sigma_e$ if estimates of the size and stellar mass are available (i.e. using a `Fundamental Plane' approach).

\begin{table*}
 \centering
 \caption{Dependence of coefficient $k(n,t_a=R/R_e)$ in equation~(\ref{Mdyn}) which transforms the observed half-light radius and light weighted projected velocity dispersion measured within an aperture $R$ to a mass.}
 \begin{tabular}{ccccccc}
 \hline
  $n$ & $t_{a} = 0.1$ & $t_{a} = 0.125$ & $t_{a} = 0.25$ & $t_{a} = 0.5$ & $t_{a} = 0.75$ & $t_{a} = 1$\\
 \hline
$ 2.00$ & $   7.38$ & $   7.20$ & $   6.80$ & $   6.78$ & $   6.97$ & $   7.30$ \\
$ 2.50$ & $   6.59$ & $   6.46$ & $   6.23$ & $   6.36$ & $   6.63$ & $   6.97$ \\
$ 3.00$ & $   5.84$ & $   5.76$ & $   5.69$ & $   5.96$ & $   6.27$ & $   6.62$ \\
$ 3.50$ & $   5.18$ & $   5.15$ & $   5.21$ & $   5.57$ & $   5.92$ & $   6.27$ \\
$ 4.00$ & $   4.62$ & $   4.62$ & $   4.79$ & $   5.21$ & $   5.58$ & $   5.93$ \\
$ 4.50$ & $   4.14$ & $   4.17$ & $   4.42$ & $   4.88$ & $   5.26$ & $   5.60$ \\
$ 5.00$ & $   3.74$ & $   3.79$ & $   4.09$ & $   4.58$ & $   4.95$ & $   5.29$ \\
$ 5.50$ & $   3.39$ & $   3.46$ & $   3.79$ & $   4.29$ & $   4.67$ & $   4.99$ \\
$ 6.00$ & $   3.10$ & $   3.17$ & $   3.52$ & $   4.03$ & $   4.40$ & $   4.71$ \\
$ 6.50$ & $   2.84$ & $   2.92$ & $   3.28$ & $   3.78$ & $   4.14$ & $   4.44$ \\
$ 7.00$ & $   2.61$ & $   2.70$ & $   3.06$ & $   3.56$ & $   3.91$ & $   4.19$ \\
$ 7.50$ & $   2.41$ & $   2.50$ & $   2.86$ & $   3.35$ & $   3.68$ & $   3.95$ \\
$ 8.00$ & $   2.23$ & $   2.32$ & $   2.68$ & $   3.15$ & $   3.47$ & $   3.73$ \\
$ 8.50$ & $   2.07$ & $   2.16$ & $   2.51$ & $   2.96$ & $   3.27$ & $   3.52$ \\
$ 9.00$ & $   1.92$ & $   2.01$ & $   2.36$ & $   2.79$ & $   3.08$ & $   3.32$ \\
$ 9.50$ & $   1.79$ & $   1.88$ & $   2.21$ & $   2.63$ & $   2.91$ & $   3.13$ \\
$10.00$ & $   1.67$ & $   1.75$ & $   2.08$ & $   2.48$ & $   2.74$ & $   2.95$ \\
 \hline
 \end{tabular}
 \label{tab:kn}
\end{table*}

\section{Dynamical mass estimates}\label{sec:Mdyn}

\subsection{Using the shape of the observed light profile: M$^{\rm PS}$}
If the contribution of dark matter to $\sigma_{\rm R}$ is negligible, the galaxy is not rotating and has an isotropic velocity dispersion, and the total mass-to-light ratio is constant, then Jeans' equation implies that the shape of the luminosity weighted projected velocity dispersion profile is fully determined by the shape of the surface-brightness profile.  Figure~14 of (the Jeans equation analysis of) Prugniel \& Simien (1997) shows $\sigma_R$ if the surface brightness profile is S{\'e}rsic with index $n$ and projected half-light radius $R_e$.  Therefore, one can estimate the dynamical mass by finding that mass-to-light ratio which correctly predicts the amplitude of $\sigma_R$ on one scale; if all the assumptions just stated are accurate, then this same value will be returned whatever $R$ one chooses to match.

\begin{figure}
 \centering
 \includegraphics[width=8cm]{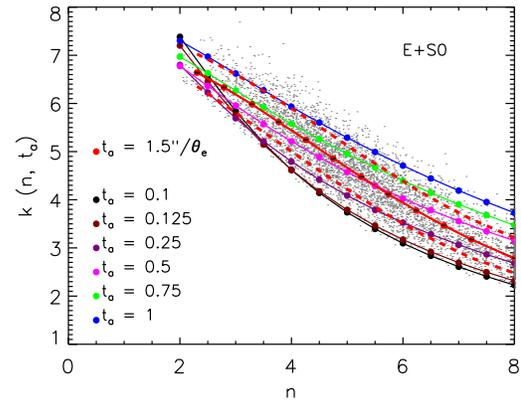}
 \caption{Proportionality constant $k(n,t_a)$ in equation~(\ref{Mdyn}) for a range of choices of $t_a$ as labeled.  Small grey dots show the values associated with the velocity dispersion measured within the SDSS fiber (i.e. $1.5$~arcsec) of $5000$ randomly selected galaxies; solid red and dashed curves show the median value, and the range which encloses 68\% of the galaxies at each $n$.  Other colored dots connected by lines show curves of fixed $t_a$ as labeled.}
 \label{knSDSS}
\end{figure}

\begin{table*}
 \centering
 \caption{Four determinations of the parameter values for equation~(\ref{Mmanga}), the correlation between $M_*^{\rm JAM}/M_*-\sigma_e$ from the literature.  We have added 0.25~dex to the values of $a$ because we use a Chabrier IMF as the fiducial value whereas the literature used Salpeter.}
 \begin{tabular}{lccc}
 \hline
 Source & $a$ & $b$ & $\Delta_{\rm rms}$ \\
  \hline
 Li et al. 2017 (E pPXF) & $-1.149\pm0.005$ & $0.591\pm0.030$ & $0.063$\\
 Li et al. 2017 (E STARLIGHT) & $-0.836\pm0.006$ & $0.457\pm0.033$ & $0.082$\\
 ATLAS$^{\rm 3D}$ (M$^{{\rm JAM}} < 2 \times 10^{11}$ M$_{\odot}$) &  $-0.621\pm0.008$ & $0.353\pm0.059$ & $0.120$\\
 ATLAS$^{\rm 3D}$ (All) &  $-0.419\pm0.008$ & $0.258\pm0.052$ & $0.120$\\
 \hline
 \end{tabular}
 \label{tab:IMF}
\end{table*}

Namely, the total mass can be estimated using
\begin{equation}
  M^{{\rm PS}(R)}\equiv k(n,R)\, R_e\,\sigma_R^2/G,
 \label{Mdyn}
\end{equation}
where $k(n,R)$ is determined using the Prugniel-Simien methodology for each $R$.  Table~\ref{tab:kn} gives these values for $n=[2,10]$ and a variety of choices of $R/R_e$.  (These values are consistent with Table~4 of Prugniel \& Simien 1997 for $R=R_e/10$, and equation~(20) of Cappellari et al. 2006 for $R=R_e$.  For this we do not use truncated profiles; doing so would make $k$ about 10 percent smaller.)  If the assumptions above are accurate, then $k(n,R)\sigma_R^2$, and hence $M^{{\rm PS}(R)}$, will be the same for all $R$.  However, if the observed scale dependence of $\sigma_R$ is different from that predicted -- e.g. if dark matter contributes to $\sigma_R$ -- then equation~(\ref{Mdyn}) will yield estimates $M^{{\rm PS}(R)}$ that depend on $R$.

To illustrate, the ratio of the mass estimates on the scales $R_e$ and $R_e/8$ will satisfy 
\begin{equation}
  \frac{M^{{\rm PS}(e/8)}}{M^{{\rm PS}(e)}} = \frac{k(n,R_e/8)}{k(n,R_e)}\,
                       \frac{\sigma_{e/8}^2}{\sigma_{e}^2}.
\end{equation}
If the assumptions above are accurate, then $M^{{\rm PS}(e/8)}=M^{{\rm PS}(e)}$.  However, if we use equation~(\ref{jfk}) to relate the two velocity dispersions, then the expression above becomes
\begin{equation}
  \frac{M^{{\rm PS}(e/8)}}{M^{{\rm PS}(e)}} = 1.25\, \frac{k(n,R_e/8)}{k(n,R_e)}.
  \label{Me10}
\end{equation}
For $n=(4,6,8)$ the values in Table~\ref{tab:kn} imply that $M^{{\rm PS}(e/8)}/M^{{\rm PS}(e)} = (0.97,0.84,0.78)$.

The fact that $M^{{\rm PS}(e/8)}\ne M^{{\rm PS}(e)}$ implies that the model is unrealistic:  E.g., gradients in the stellar mass-to-light ratio or the presence of dark matter would both invalidate the assumption that the total mass-to-light ratio is constant; stellar orbits may not be isotropic; etc.  For large $n$, the fact that $M^{{\rm PS}(e/8)}$ is the smaller of the two is usually viewed as indicating that dark matter is a smaller fraction of the total mass on small scales (i.e., $\sigma_e$ is more contaminated by dark matter than is $\sigma_{e/8}$).

\begin{figure}
 \centering
 \includegraphics[width=8cm]{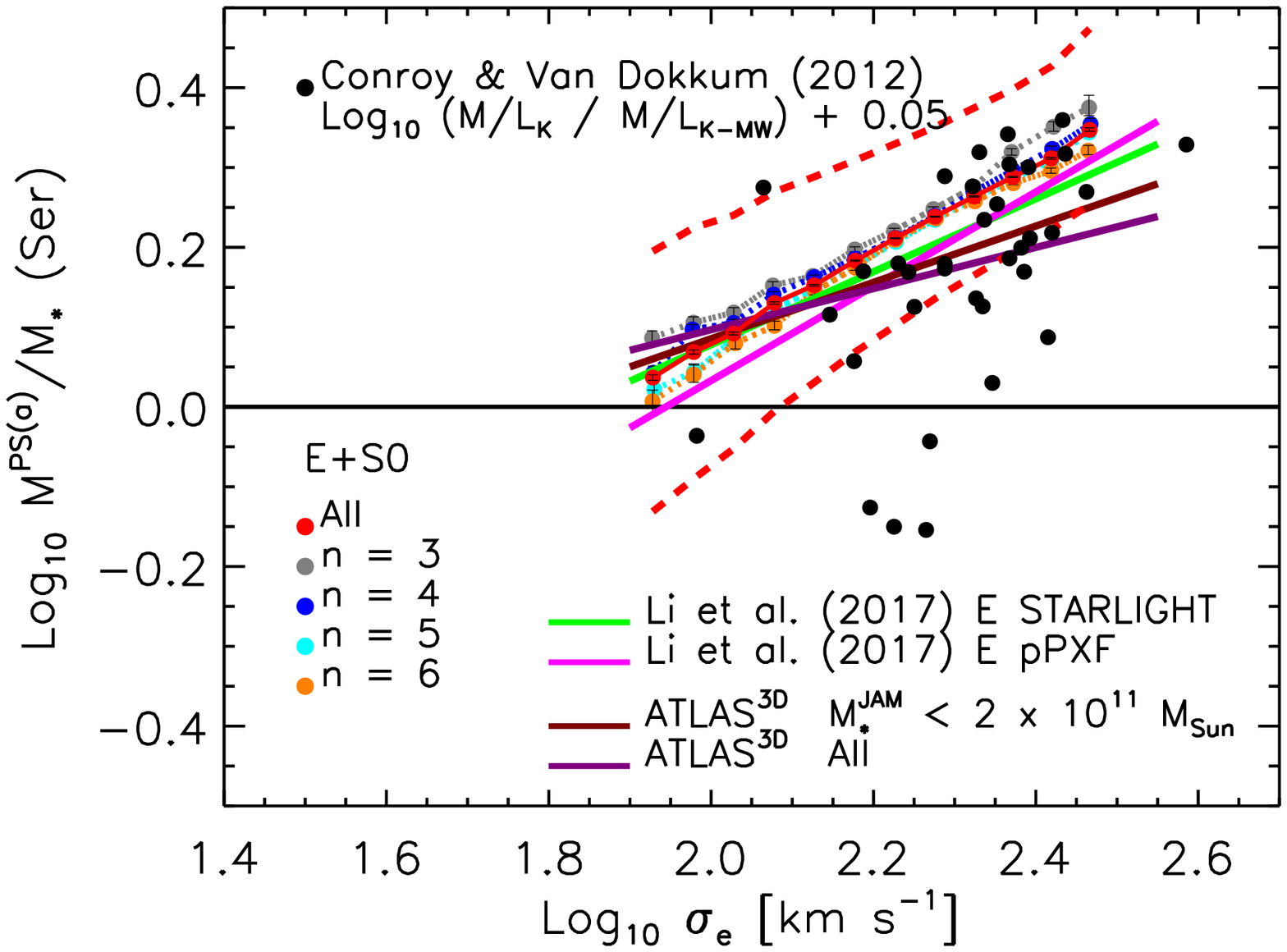}
 \includegraphics[width=8cm]{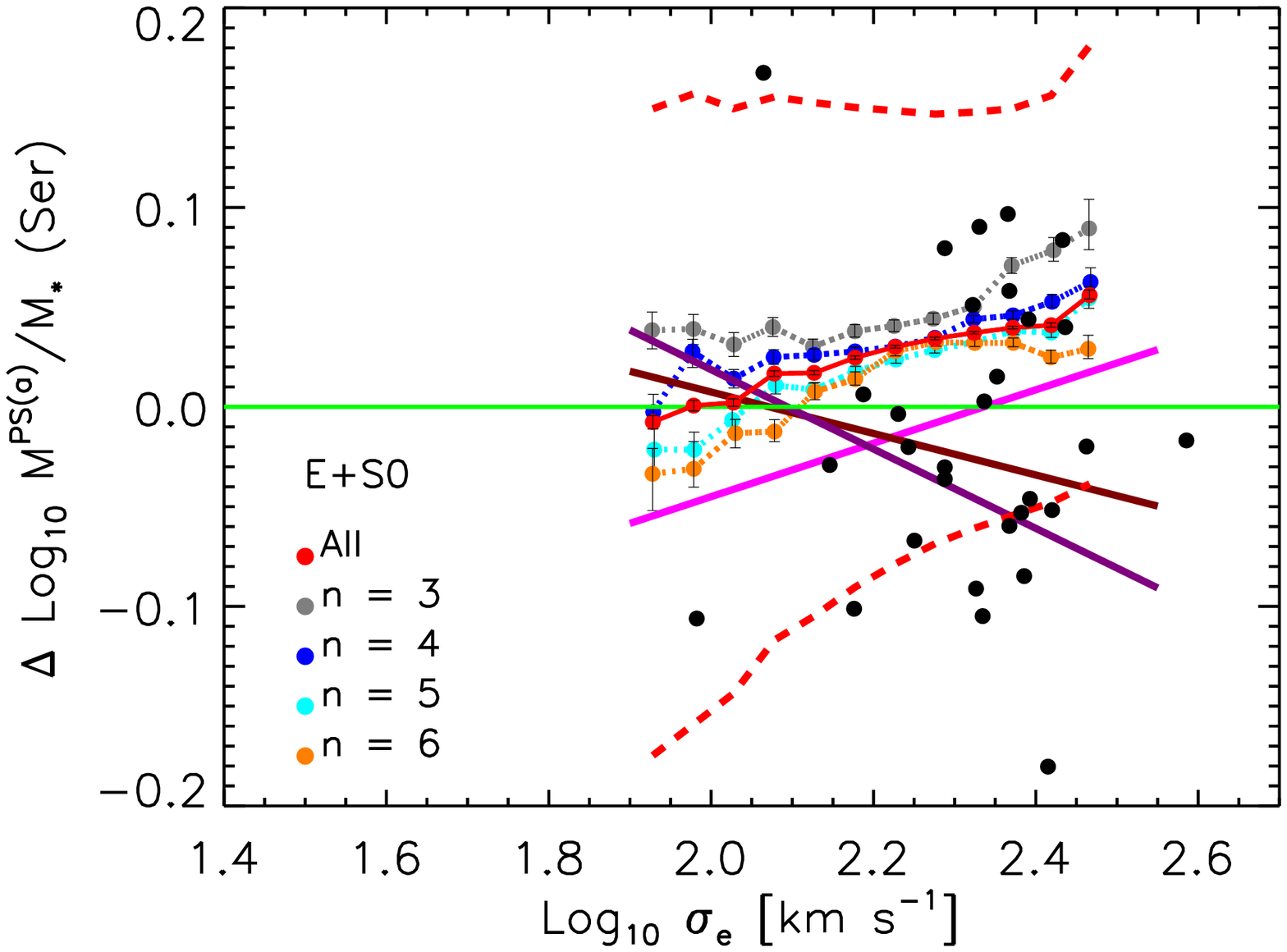}
 \caption{Top:  Ratio of the mass estimate $M^{{\rm PS}(a)}$ from equation~(\ref{Mdyn}), which was estimated from the velocity dispersion within the SDSS fiber $\sigma_a$, to the stellar mass $M_*$ based on $M_*/L$ values taken from Mendel et al. (2014) (Chabrier IMF) and truncated S{\'e}rsic luminosities, shown as a function of $\sigma_e$, for SDSS E+S0s. Red solid and dashed curves show the median value  and the range which encloses 68\% of the objects in each bin in $\sigma_e$; grey, blue, cyan and orange curves show subsamples of fixed $n$.  Green and magenta lines show the relations reported by Li et al. (2017); purple and brown curves show ATLAS$^{\rm 3D}$ scalings (equation~\ref{Mmanga} and Table~\ref{tab:IMF}).  Filled black symbols show the objects studied by Conroy \& van Dokkum (2012; see text for details).  Bottom:  Lines and symbols as top, except that now residuals with respect to the green line labeled {\tt E STARLIGHT} are shown. }
 \label{M*a}
\end{figure}

\begin{figure}
 \centering
 \includegraphics[width=8cm]{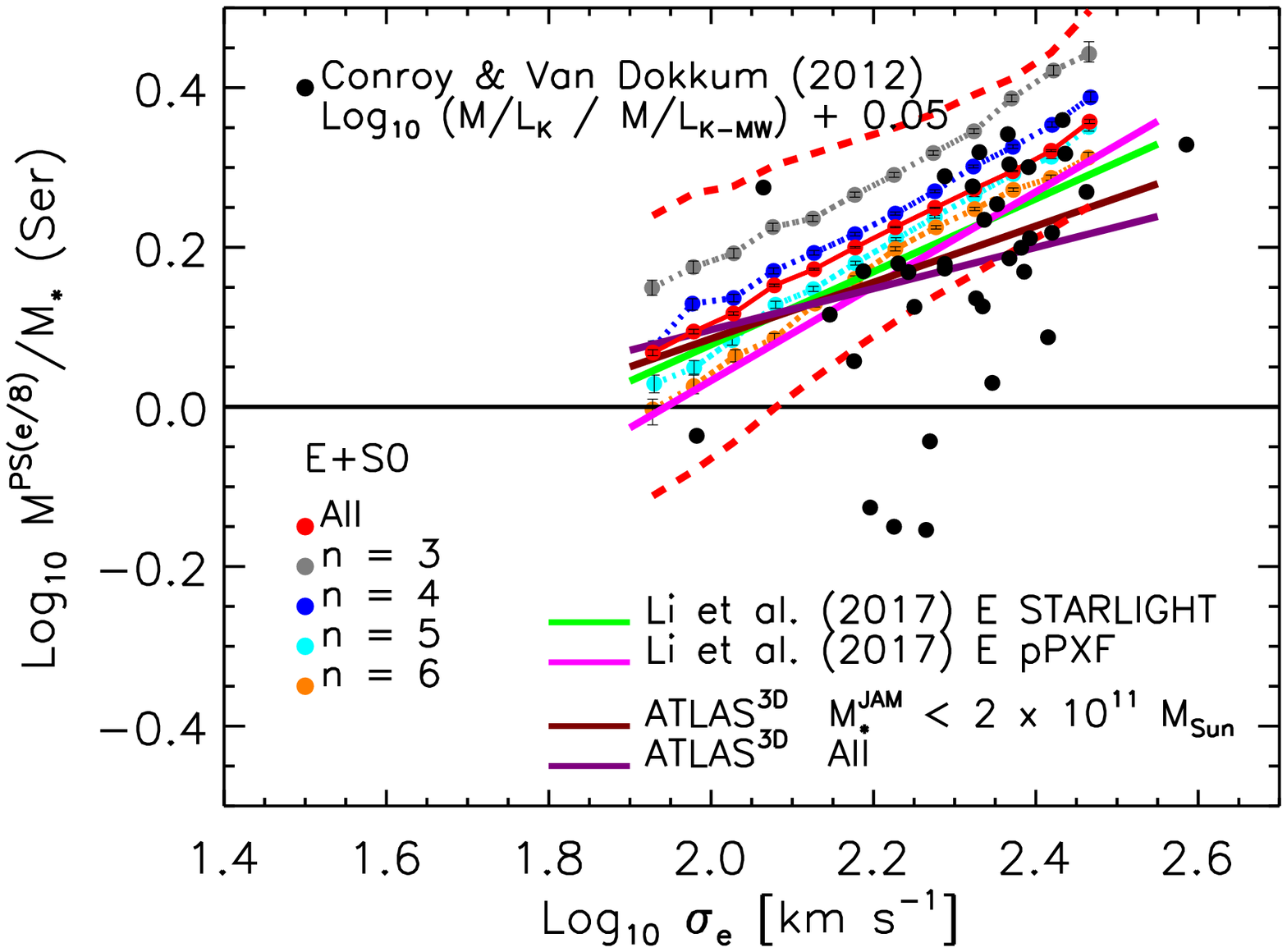}
 \includegraphics[width=8cm]{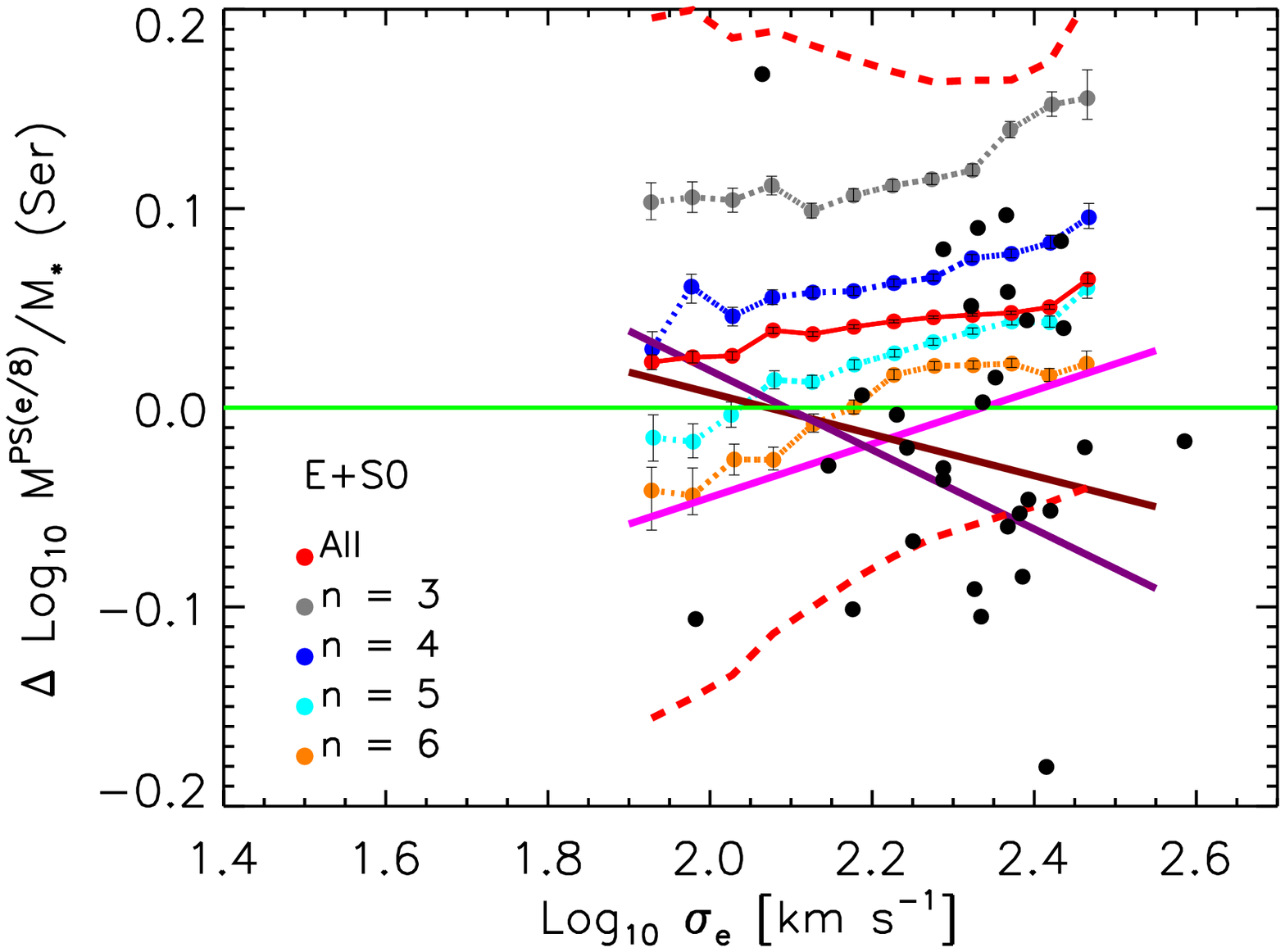}
 \caption{Same as previous figure, but now when equation~(\ref{Mdyn}) is normalized to match $\sigma_{e/8}$.  The dependence on $n$ is stronger than in the previous Figure.}
 \label{M*e8}
\end{figure}

\subsection{Simple analysis of E+S0s}\label{eso}
With this in mind, we have selected the subsample of objects in the SDSS for which the S{\'e}rsic index $n\ge 2$ and the BAC probability $p$(E+S0) $\ge 0.7$, where $p$(E+S0) is the sum of the BAC $p$(E) and $p$(S0) values.  These are the objects for which $M_*^{\rm PS}$ has the best chance of being accurate, and we refer to them as the SDSS E+S0s.  In many of the plots which follow, smooth (usually red) solid curve shows the median of the ordinate for narrow bins in the abscissa, and dashed curves show the range which encloses 68\% of the objects.

Following the discussion of the previous subsection, we produced two simple estimates of the stellar dynamical mass of these E+S0s:  $M_*^{{\rm PS}(a)}$ and $M_*^{{\rm PS}(e/8)}$.  The former uses the measured $\sigma_a$, i.e. the velocity dispersion measured within the SDSS fiber ($1.5$~arcsec), and the other uses $\sigma_{e/8}$, obtained by extrapolating $\sigma_a$ using equation~(\ref{jfk}).  Since $R_a/R_e$ is different for each object, there is a distribution of $k(n,t_a)$ values at each $n$ (which are needed to estimate $M_*^{{\rm PS}(a)}$); the small dots in Figure~\ref{knSDSS} show this distribution.  The other smooth curves show the relations given in Table~\ref{tab:kn}; the brown curve for $k(n,t_a=0.125)$ is what we use when defining $M_*^{{\rm PS}(e/8)}$.

Since $R_a\approx 0.6R_e$ on average, we expect that $M_a$ is more likely to have been contaminated by dark matter; the literature suggests that the dark matter fraction within $R_e$ is of order 15\% \cite{tiret10,atlas3dXX,chae14,shankar17}.  Within the scale associated with the SDSS fiber aperture $\theta_a\approx\theta_e/2$ this fraction is about $2\times$ smaller (e.g. Figure~9 in Shankar et al. 2017), and it is smaller still within $R_{e/8}$.  (These estimates depend somewhat on the model for the dark matter, but there is general agreement that the fraction decreases on smaller scales.)

The top panel of Figure~\ref{M*a} shows $M_*^{\rm PS(a)}/M_*$ for E+S0s as a function of $\sigma_e$.  We have chosen this format for ease of comparison with the work of Li et al. (2017).  For $M_*$, we have multiplied the truncated S{\'e}rsic $L_{\rm trunc}$ by $M_*/L$ from the dust-free, Chabrier IMF models of Mendel et al. (2014, see Appendix~\ref{ml}). The red solid and dashed curves show the median $M_*^{\rm PS(a)}/M_*$ in each bin in $\sigma_e$, and the range which encloses 68\% of the objects. The other colored curves, which are almost superimposed on one another, show the result of restricting to narrow bins in S{\'e}rsic $n$.

The straight lines show relations of the form 
\begin{equation}
 \Big\langle \log_{10}\frac{M_*^{\rm JAM}}{M_*}\Big|s_e\Big\rangle = a + b\,s_e,
 \ {\rm where} \ s_e\equiv \log_{10}\frac{\sigma_e}{\rm km~s^{-1}},
 \label{Mmanga}
\end{equation}
and the coefficients $a$ and $b$, provided in Table~\ref{tab:IMF}, are taken from the recent literature \cite{atlas3dXV,mangaIMF}.  (We have shifted the zero-point $a$ from the literature, where Salpeter was the fiducial choice, by 0.25~dex so that it conforms to our choice of a Chabrier IMF as the fiducial value.)  The larger black circles show the objects studied by Conroy \& van Dokkum (2012); in this case the y-axis shows their estimate of the ratio of the variable and fixed IMF $M_*$ values.  (We have shifted their $M_*$ estimate to account for the fact their Milky Way IMF is based on Kroupa 2001, which differs by 0.05~dex from Chabrier 2003.)  Neither their quantity nor the straight lines are explicitly $M_*^{\rm PS}/M_*$, yet the agreement with our SDSS estimates is remarkably good.

The bottom panel shows residuals with respect to the line labeled {\tt E STARLIGHT}: equation~(\ref{Mmanga}) with $(a,b) = (-0.836,0.457)$.
This removes most of the trend with $\sigma_e$ (note the range along the y-axis, is now $2.5\times$ smaller than in the top panel), and highlights the weak trend with $n$. Correcting $M_*^{\rm PS}$ downwards by $\sim 0.07$~dex to account for dark matter would bring our estimates into quite good agreement with the {\tt E pPXF} relation (magenta).  On the other hand, the rms scatter is larger than the values reported in Table~\ref{tab:IMF}.  

Figure~\ref{M*e8} shows a similar analysis, but now from inserting $\sigma_{e/8}$, which was obtained from the measured $\sigma_a$ using equation~(\ref{jfk}), in equation~(\ref{Mdyn}).  Although the median is again in good agreement with equation~(\ref{Mmanga}), the dependence on $n$ is stronger.  This is because $M_{e/8}<M_a$ at $n>4$, whereas the trend is reversed at smaller $n$. Although the agreement between these estimates and the literature is already quite good, the next subsection asks if it can be improved further.

\begin{figure}
 \centering
 \includegraphics[width=8cm]{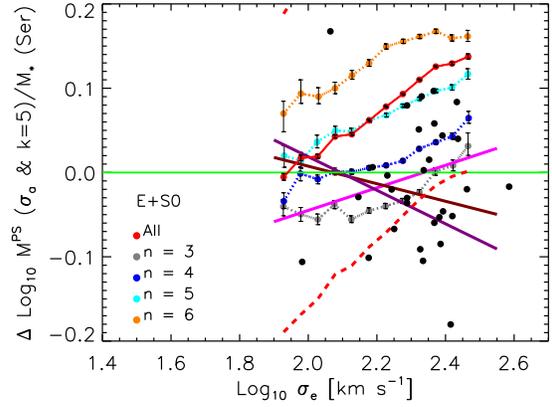}
 \caption{Same as bottom panel of Figure~\ref{M*a}, but now with $k$ in equation~(\ref{Mdyn}) fixed to $5$.  Ignoring the $n$ dependence of $k$ biases the median relation and makes the scatter around it larger.}
 \label{k5}
\end{figure}

Before we do so, however, it is useful to illustrate the importance of the $k$-dependence on $n$ when computing $M^{{\rm PS}}$ if the half-light radius $R_e$ is from a S{\'e}rsic fit to the light profile. Figure~\ref{k5} uses $5\,R_e\sigma_a^2/G$ in place of $M^{\rm PS}$ (i.e. the $n$ dependence of $k$ is ignored).  This should be reasonably accurate for galaxies with $n\approx 4.5$, but not otherwise (c.f. Table~\ref{knSDSS}), and indeed, figure shows a strong dependence on $n$.  Using a different $k$-independent value for $k$ would shift all the points up or down, but would not remove the $n$-dependence.  Using $k(n)$, as was done in the bottom panel of Figure~\ref{M*a}, reduces the $n$-dependence substantially.  For this reason, and because there is no physical motivation for using the same $k$ for all $n$, we do not consider the fixed-$k$ estimate further.  

\subsection{Calibration using $M_*^{\rm JAM}$ of ATLAS$^{\rm 3D}$}
\begin{figure}
 \centering
 \includegraphics[width=8cm]{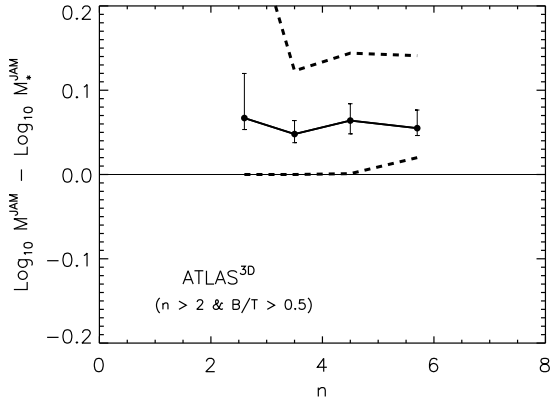}
 \caption{Ratio of the total to stellar mass estimate from JAM (Cappellari et al. 2013b).}
 \label{fDMa3d}
\end{figure}

Most of this subsection is devoted to using the ATLAS$^{\rm 3D}$ dataset (Cappellari et al. 2011) to calibrate our mass estimates for SDSS galaxies.  This considerably smaller sample has considerably richer data -- spatially resolved spectra -- for each object in it.  This enabled modelling which allows for anisotropic velocity dispersions as well as a dark matter component.  On the basis of such models, Cappellari et al. (2013b) argue that $M^{{\rm PS}(e)}$ over-estimates their best estimate of the total mass, which they call $M^{\rm JAM}$.  This total mass is the sum of the mass in stars $M_*^{\rm JAM}$, gas and dark matter.  The bottom right panel of their Figure~14 shows that $M^{\rm JAM}\approx 0.7\,M^{{\rm PS}(e)}$.  Their estimate of the red stellar component, $M_*^{\rm JAM}$, is smaller by a factor of $(1-f_{\rm DM})$ with $f_{\rm DM}\approx 0.12$ estimated within a sphere of radius $r=R_{e\rm MGE}$ \cite{atlas3dXX}.  Figure~\ref{fDMa3d} shows the ratio of their total and stellar mass estimates for a subsample of 86 galaxies which have $n>2$ and bulge disk decompositions with bulge-to-total ratios $B/T > 0.5$, which should better represent our SDSS E+S0s sample.  Our results do not change significantly if we change this selection.  

The analysis of Section~3.1 suggests that at least some of this difference is because they chose to normalize the simple PS model to $\sigma_{e}$, the velocity dispersion observed on the scale $R_{e}$.  Had they chosen to work with $M^{{\rm PS}(e/8)}$ rather than $M^{{\rm PS}(e)}$, then the required correction factor -- which accounts for at least some of the contribution from dark matter -- would have been closer to unity, at least when $n\ge 4$.  

\begin{figure}
  \centering
 \includegraphics[width=8cm]{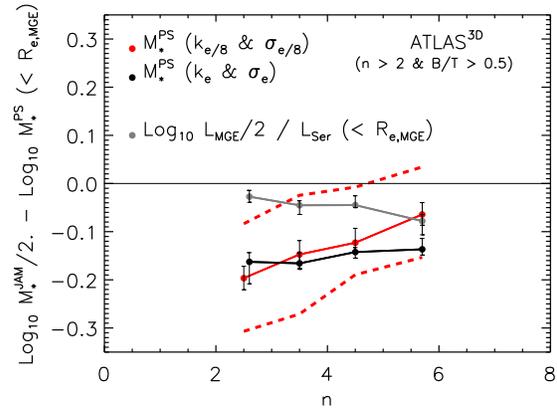}
 \caption{Ratio of $M_*^{\rm JAM}/2$, the stellar mass estimate within the projected half-light radius $R_{e\rm MGE}$ from JAM, to that from the S{\'e}rsic-profile based equation~(\ref{Mdyn}), shown as a function of S{\'e}rsic index $n$.  Results for two choices of the scale $R$ on which $M^{{\rm PS}(R)}$ was normalized are shown:  $R=R_{e\rm MGE}$ (black) and $R_{e\rm MGE}/8$ (red).  Grey symbols show a corresponding analysis of the light within $R_{e\rm MGE}$. The $M^{{\rm PS}(R)}$ estimate with $R=R_{e\rm MGE}/8$ is in better agreement with the JAM estimate, and even more so once accounting for the difference in luminosity.}
 \label{redoa3d}
\end{figure}

Figure~\ref{redoa3d} shows the result of re-analysing their data in this way: the black line shows the median value of their estimate of the stellar mass within their estimate of the projected half light radius, $R_{e\rm MGE}$ (the parameters were taken from Table~1 of Cappellari et al. 2013a and~Cappellari et al. 2013b, i.e. ATLAS$^{\rm 3D}$ Papers XV and XX), divided by the fraction of the S{\'e}rsic-based estimate 
 $M_{e\rm MGE} = k(n,R_{e\rm MGE})\,R_{e\rm Ser}\,\sigma^2_{e\rm MGE}/G$
within the same radius, plotted as a function of $n$.  The S{\'e}rsic-based quantities are taken from Table~C1 of Krajnovic et al. 2013 (ATLAS$^{\rm 3D}$ Paper XVII; when transforming their apparent to absolute magnitudes, we multiply their $L$ estimate by their bulge-flattening parameter $q_B$).  Note that $R_{e\rm MGE}<R_{e\rm Ser}$ because $L_{\rm Ser}$ is larger than their fiducial $L$, so the relevant value for $k$ is not quite that for $t_a=1$.


The black line should be compared with the red solid curve, for which the S{\'e}rsic-based estimate uses $\sigma_{e{\rm MGE}/8}$ and $k(n,R_{e\rm MGE}/8)$ (so $k$ is close to, but not quite that for $t_a=1/8$).  The figure shows that the $M_{e{\rm MGE}/8}$-based estimate is closer to $M_*^{\rm JAM}$, presumably because normalizing the model to $\sigma$ on a smaller scale has provided an estimate which is less contaminated by dark matter.  We have checked that repeating the analysis within a sphere of radius $r_{1/2}$ (a scale which they also provide), yields similar results.

The remaining discrepancy depends on $n$.  
While this may indicate that dark matter or anisotropic velocity dispersions still matter on the scale $R_{e\rm MGE}/8$ (but note that the median value of $R_{e\rm MGE}/8$ for $n\ge 4$ is of order 500~pc), some of it may be due to systematics.  To motivate this suggestion, the grey symbols show a similar analysis of their estimates of the projected light within the same radius (from their Papers XV and XVII).  Clearly, the MGE and S{\'e}rsic estimates of the light within $R_{e\rm MGE}$ are not the same. Therefore, it is the difference from the grey curve in Figure~\ref{redoa3d} that is likely to be a better measure of the discrepancy between the simpler PS and more sophisticated JAM estimates.  For the red curve, this difference is small at large $n$, but is larger as $n$ decreases.  For the black, the dependence on $n$ is much weaker.

\subsection{Final calibrated estimate: $M_*^{\rm PS-JAM}$}
If we assume that the JAM estimates of the stellar mass are accurate, then we must correct the S{\'e}rsic-based estimates (e.g. Figure~\ref{redoa3d}) before using them.  Our goal is to define correction factors which can be applied to the SDSS dataset.

With this in mind, we note that, at $n\ge 4$, the typical value of $R_{e\rm MGE}/R_e$ in the bulge dominated ATLAS$^{\rm 3D}$ sample is $\sim 0.8$.  In the SDSS, $R_a\sim 0.6R_e$, where $R_a$ is the aperture radius of the fiber within which $\sigma_{\rm a}$ was estimated.  This makes $R_{e\rm MGE}/8\approx 0.8R_e/8 = 0.1R_e$.  So, to apply our $R_{e\rm MGE}/8$ results to the SDSS, we must extrapolate the measured SDSS $\sigma_a$ to $\sim R_e/10$. In practice, the difference between $\sim R_e/10$ and $\sim R_e/8$ is negligible (see Figure~\ref{knSDSS}), so we use $\sim R_e/8$.  We describe this aperture correction before moving on to describe our correction factors.

\subsubsection{Calibrating the velocity dispersion profiles}\label{sigcorr}
We noted above that, if we would like to work with $\sigma_{e/8}$, then we must apply an aperture correction to the measured SDSS values.  (In contrast, in ATLAS$^{\rm 3D}$, $\sigma_{e\rm MGE/8}$ was measured, not extrapolated.)  Whereas correcting using equation~(\ref{jfk}) is standard, it assumes that the profile shape is the same for all galaxies.  However, the observed scatter in $\sigma$ profiles is large \cite{sauronIV}, so using an average scaling to aperture correct to different scales may introduce systematic errors.  Moreover, the mean correction may, in fact, depend on the shape of the light profile.  Therefore, we have considered the possibility that
 \begin{equation}
  {\rm log}_{10}\frac{\sigma_R}{\sigma_e}
  = -\gamma(n,R,R_e)\, {\rm log}_{10} \frac{R}{R_e} ,
 \label{gamman}
\end{equation}
where $n$ is the S{\'e}rsic index.

\begin{figure}
 \centering
 \includegraphics[width=8cm]{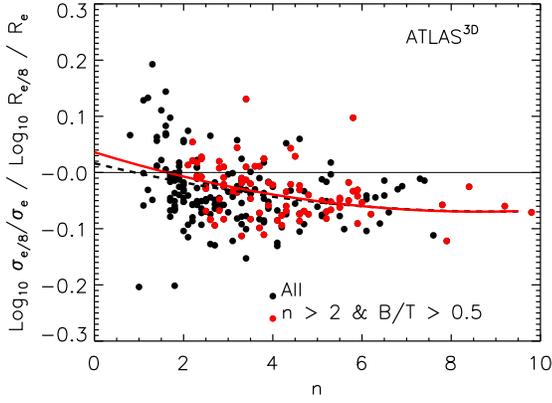}
 \caption{Ratio of the velocity dispersion measured within an aperture of size on $R_{e\rm MGE}/8$ to that within $R_{e\rm MGE}$, as a function of S{\'e}rsic index $n$, for all galaxies in the ATLAS$^{\rm 3D}$ survey (black) and the subset that have $n>2$ and B/T$>0.5$ (red).  This ratio is usually assumed to equal $-0.066$ for all $n$.  Dashed and solid lines show equation~(\ref{gamman}) with the $n$-dependence of equations~(\ref{gammaAll}) and~(\ref{gammaSub}). 
 }
 \label{sigma3d}
\end{figure}

Figure~\ref{sigma3d} shows $-\gamma(n,R_{e\rm MGE/8},R_{e\rm MGE})$ for the galaxies in the ATLAS$^{\rm 3D}$ survey, taken from Table~1 of Cappellari et al. (2013a) and~Cappellari et al. (2013b), plotted as a function of $n$ (taken from Table~C1 of Krajnovic et al. 2013).  Filled black circles show all objects (`All' in the expression below) and red show the subset with $n>2$ and B/T$>0.5$ (recall that these are most like the E+S0 SDSS sample).  We refer to these as `All' and 'Sub' in the expressions below.  In both cases there is a wide range of slopes $\gamma$ at each $n$, and there is a weak tendency for $\gamma$ to be closer to zero as $n$ decreases.  Lines show quadratic fits to these samples:  
\begin{align}  
  \gamma_{\rm All} &= a_{\rm All} + b_{\rm All}\,(n-4) + c_{\rm All}\, (n-4)^2;
\label{gammaAll}\\
  \gamma_{\rm Sub} &= a_{\rm Sub} + b_{\rm Sub}\,(n-4) + c_{\rm Sub}\, (n-4)^2,
 \label{gammaSub}
\end{align}
with ($a_{\rm All}$, $b_{\rm All}$,  $c_{\rm All}$) $=$ ($0.0431\pm 0.0042$, $0.0106\pm 0.0021$, $-0.0011\pm 0.0005$) and ($a_{\rm Sub}$, $b_{\rm Sub}$,  $c_{\rm Sub}$) $=$ ($0.0392\pm 0.0045$, $0.0132\pm 0.0035$, $-0.0014\pm  0.0007$).

The relatively small size of the ATLAS$^{\rm 3D}$ sample means that while this trend with $n$ is clear, quantifying it precisely requires a bigger sample.  Nevertheless, we have found that if we wish to predict the red curve shown in Figure~\ref{redoa3d} by aperture correcting the measured $\sigma_{e\rm MGE}$ to $\sigma_{e\rm MGE}/8$ (rather than measuring it directly), then this $n$-dependent correction is slightly more accurate than is equation~(\ref{jfk}).  To see why, note that if we aperture correct using equation~(\ref{gamman}), then 
\begin{equation}
  \frac{M^{{\rm PS}(e/8)}}{M^{{\rm PS}(e)}} =
  8^{2\gamma(n)}\, \frac{k(n,R_e/8)}{k(n,R_e)},
\end{equation}
making $M^{{\rm PS}(e/8)}/M^{{\rm PS}(e)}$ differ from equation~(\ref{Me10}) by more than 10\% at $n\le 4$.  This is relevant because, in the SDSS, we only have velocity dispersion measurements on the scale of the SDSS fiber.  So if we wish to work with $M^{{\rm PS}(e/8)}$ in the SDSS, then estimating $\sigma_{e/8}$ from the measured $\sigma_a$ is an important step.

\subsubsection{Correcting PS to JAM}
When we return to work with the SDSS dataset, we will actually work with 
\begin{align}
  \label{Mfinal}
  M_{*}^{\rm PS-JAM} &\equiv (1 - f_{\rm gas})\,C_R(n)\, M^{{\rm PS}(R)}\nonumber\\
                  &= (1 - f_{\rm gas})\,\kappa(n, R)\, R_e\sigma^2_R/G.
\end{align}
Here $f_{\rm gas}$ corrects for the gas fraction, 
$C_R(n)$ corrects for the analog of the difference between the red and grey curves in Figure~\ref{redoa3d}, and we have defined $\kappa(n,R)\equiv C_R(n)\,k(n,R)$ as an overall $n$- and $R$-dependent correction factor.

The first term, $1-f_{\rm gas}$, is necessary because it is common to quote $M_*$ estimates which ignore the contribution from gas (which is usually implicitly assumed to have the same spatial distribution as the stars).  For E+S0s, we set $f_{\rm gas}=0$, since this was done by Li et al. (2017), with whose work we compare in the next section.  For spirals, their Figure~2 suggests that $f_{\rm gas}\approx \exp[-(M_*/10^9\, M_\odot)]$, which we use in what follows.

\begin{figure}
 \centering
 \includegraphics[width=8cm]{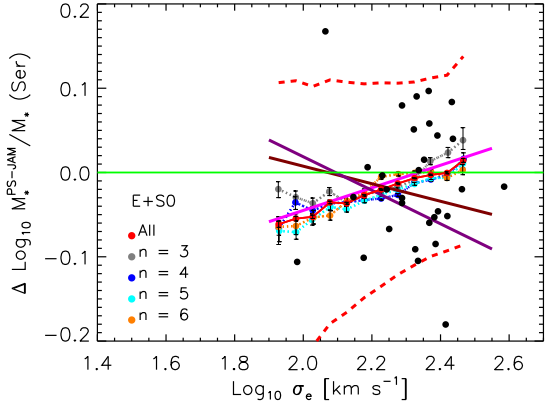}
 \caption{Same as bottom panels of Figures~\ref{M*a} and~\ref{M*e8} (i.e. difference with respect to the E STARLIGHT fit of Table~\ref{tab:IMF}), except that the numerator is from equation~(\ref{Mfinal}). The magenta, brown and purple lines are the same as in Figure~\ref{M*a}. Calibrating to JAM (using equations~\ref{Fjam8} and~\ref{FjamR}) has removed the $n$-dependence (compare Figures~\ref{M*a} and~\ref{M*e8}).}
 \label{M*final}
\end{figure}

\begin{figure}
 \centering
 \includegraphics[width=8cm]{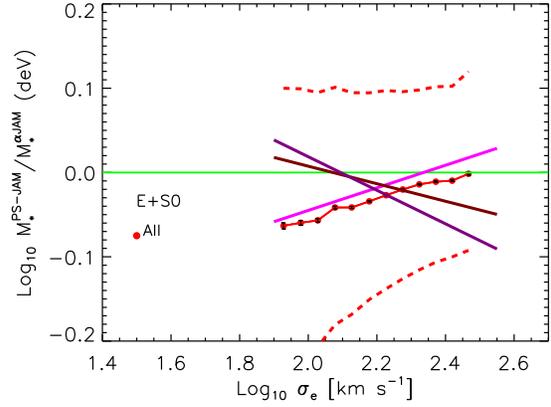}
 \caption{Same as Figure~\ref{M*final}, except that now $M_{*}^{\rm PS}$ and $L_{\rm trunc}$ are determined from de Vaucouleurs fits to the light profiles. The magenta, brown and purple lines are the same as in the previous Figure.}
 \label{M*deV}
\end{figure}

The other correction factor, $C_R(n)$, depends on the scale $R$ on which the $\sigma$ used for estimating $M^{{\rm PS}(R)}$ was measured (because the black and red curves in Figure~\ref{redoa3d} are different from one another).  
For $R_e/8\approx R_{e\rm MGE/8}$, the correction required to bring the red curve to the grey is reasonably well approximated by 
\begin{equation}
  C_{e/8} = \frac{M_*^{\rm JAM}}{M_{e\rm MGE/8}}
         \approx 10^{-0.054(\pm 0.011) + 0.022(\pm 0.008)\,(n-4)}
  \label{Fjam8}
\end{equation}
over the range $2\le n\le 6$.  For $n<2$ and $n>6$ we set $C$ equal to its value at $n=2$ and $n=6$ respectively.  (While this is formally not the best-fitting linear relation, it is within 1$\sigma$ of the best-fit, chosen because it gives slightly cleaner, i.e., less $n$-dependent, results in Figures~\ref{M*final} and~\ref{M*deV}.)  

For other $R$, we use the fact that 
\begin{equation}
 C_R = C_{e/8}\,\frac{M^{{\rm PS}(R_e/8)}}{M^{{\rm PS}(R)}} 
     = C_{e/8}\,\frac{k(n,R_{e/8})}{k(n,R)}\frac{\sigma^2_{e/8}}{\sigma_R^2},
 \label{FjamR}
\end{equation}
with equation~(\ref{gamman}) to aperture correct the velocity dispersion from one scale to another if direct measurements are not available.  This ensures that 
$M_*^{\rm PS-JAM}$ estimated from $\sigma_a$ will be the same as that estimated from $\sigma_{e/8}$.  Thus, to compute the `virial' mass, one can simply replace the factor $k(n,R)$ in equation~(\ref{Mdyn}) with $\kappa(n, R)$ (equation~\ref{Mfinal}):
\begin{equation}
  \label{kappa}
  \kappa(n, R) = C_{e/8}\, k(n,R_{e/8})\left(\frac{R}{R_{e/8}}\right)^{2\gamma(n)}  
\end{equation}
where $C_{e/8}$ is given by equation~(\ref{Fjam8}), $k(n,R_{e/8})$ is from Table~\ref{tab:kn} (third column, i.e. $t_a = 0.125$), and $\gamma(n)$ is given by equation~(\ref{gamman}).

We are finally ready to return to the comparison shown in the bottom panels of Figures~\ref{M*a} and~\ref{M*e8}.  As we noted there, although the median $\log_{10}(M_*^{\rm PS}/M_*) - \sigma_e$ relation was in reasonable agreement with previous JAM-work, there was a strong dependence on $n$.  The colored symbols in Figure~\ref{M*final} show
 $$\log_{10}(M_*^{\rm PS-JAM}/M_*) - 0.457\,s_e + 0.836$$
as a function of $\sigma_e$ for the SDSS E+S0s when $M_*^{\rm PS-JAM}$ was estimated from $\sigma_{e/8}$ and equation~(\ref{Fjam8}), with $\sigma_{e/8}$ being aperture corrected from $\sigma_a$ using equation~(\ref{gamman}).  The other (black) symbols and lines are the same as in the bottom panels of Figures~\ref{M*a} and~\ref{M*e8}.  In contrast to those previous figures, now subsamples of different $n$ superimpose.  In addition, with the dependence on $n$ removed, the scatter around the median relation is smaller (compare Figures~\ref{M*a} and~\ref{M*e8}).  

The median relation defined by the SDSS sample now happens to be in very good agreement with the {\tt E pPXF} curve of Li et al. (2017), but this is probably just a coincidence given the uncertainty in the equations we used to calibrate $M_*^{\rm PS-JAM}$ (due to the relatively small size of the ATLAS$^{\rm 3D}$ sample) and the fact that our $M_*$ was computed using different stellar population models. Note that we do not match the ATLAS$^{\rm 3D}$ line exactly because our correction factor was calibrated at fixed $n$ (see Figure~\ref{redoa3d}) rather than $\sigma_e$.

For completeness, and because deVaucouleurs photometry (i.e. S{\'e}rsic index $n$ is set to 4) remains a popular choice, Figure~\ref{M*deV} shows the result of using deVaucouleurs photometric parameters when estimating $M_*^{\rm PS-JAM}/M_*$.  Comparison with the bottom panel of Figure~\ref{M*final} shows that the deVaucouleurs estimate is very similar (it is only $\sim 0.02$~dex smaller).  This implies that $(M_*^{\rm PS-JAM}/L)_{\rm deV}\approx (M_*^{\rm PS-JAM}/L)_{\rm Ser}$. 
This is because the estimated mass is the product of three terms:  $\kappa(n,R)$, $R_e$ and $\sigma_R$ (equation~\ref{Mfinal}).  
If we ignore the $n$-dependence of the aperture correction, then the ratio of deVaucouleur and S{\'e}rsic mass estimates equals $[\kappa(4,e/8)/\kappa(n,e/8)]\,(R_{e\rm deV}/R_{e\rm Ser})^{1+2a_{\rm Sub}}$.
The $n$-dependence of $\kappa$ is primarily because $k$ decreases with $n$ (Table~\ref{knSDSS}).  Since $R_e$ typically increases, and $a_{\rm Sub}\ll 1$, what matters is the product $\kappa R_e$, which is a weaker function of $n$.  The net result is that $\kappa(4,e/8)R_{e\rm deV}/k(n,e/8)\,R_{e\rm Ser} < 1$.  However, $L_{\rm deV}/L_{\rm Ser}<1$ also, making $(M_*^{\rm PS-JAM}/L)_{\rm deV}\approx (M_*^{\rm PS-JAM}/L)_{\rm Ser}$.  
These cancellations show that one should not mix and match parameters from different fits: the simple virial estimator $5 R_e\sigma_e^2/G$ we mentioned in the Introduction (see also Figure~\ref{k5} and related discussion at the end of Section~\ref{eso}) is only well-motivated if the fitting procedure returns a size estimate that is close to $R_{\rm deV}$ (for which $\kappa\sim 5$).  

\section{Stellar mass proxies}\label{proxies}
This section compares three proxies for the dynamical stellar mass in the SDSS dataset.  These proxies use various combinations of $M_*$, $R_e$ and $\sigma_a$.  The first proxy is motivated by the fact that the ratio of dynamical and stellar population masses correlates with velocity dispersion (e.g. Figures~\ref{M*a} and~\ref{M*e8}, and equation~\ref{Mmanga}).  It uses $M_*$ and $\sigma_a$ aperture corrected to $\sigma_e$ (Section~\ref{sec:Ma}) to estimate $M_*^{\rm dyn}$.  The second is closely related, since it simply replaces $\sigma_e$ with $s_{\rm FP}(M_*,R_e)$ (equation~\ref{sigmaFP}), the Fundamental Plane approximation to it.  Hence, the proxy depends on $M_*$ and $R_e$ (Section~\ref{sec:fp}).  The third uses $R_e$ and $\sigma_a$ aperture corrected to $\sigma_{e/8}$:  this follows from the calibrations to JAM described in Section~\ref{sec:Md}.  All three are expected to be good approximations to the total stellar mass estimated using stellar population modelling which allow for a variable-IMF \cite{califaIMF}.

\subsection{$M_*^{\alpha_{\rm JAM}}$}\label{sec:Ma}
Equation~(\ref{Mmanga}) suggests that one can combine a fixed-IMF estimate of $M_*$ (in our case, Chabrier), with the value of $\sigma_e$ we estimate from the measured $\sigma_{\rm a}$ (for this, the choice of aperture correction makes little difference), to build a proxy for $M_*^{\rm JAM}\approx M_{*,\rm IMF}$.  We call this proxy $M_*^{\alpha_{\rm JAM}}$, where 
\begin{equation}
 \log_{10}\frac{M_*^{\alpha_{\rm JAM}}}{M_\odot} \equiv \log_{10}\frac{M_*}{M_\odot} + a + b\, s_e + \Delta_{\rm rms},
 \label{Mjam}
\end{equation}
where $(a,b)$ can be any of the pairs in our Table~\ref{tab:IMF}, and $\Delta_{\rm rms}$ is a Gaussian number with rms $\sim 0.1$~dex (Table~\ref{tab:IMF}).  (Recall that we have added 0.25~dex to the values of $a$ from the literature so that they conform to our choice of a Chabrier rather than Salpeter IMF as the fiducial value.) Normalized as we do, the correction to Chabrier is small at $\sigma_e\sim 100$~km~s$^{-1}$, but grows at larger $\sigma$.  In addition, low mass rotators (those for which $\sigma_{\rm a}$ may be contaminated by rotation) are objects for which $M_{*}^{\rm \alpha_{JAM}}\approx M_*$, so the correction is small.  In addition, as we will see shortly, at low masses $\phi(M_*)$ is flat, so the correction is unimportant.  That is to say, the objects for which our correction is least well-motivated are those for which the correction matters little.  (Note that equation~\ref{Mmanga} was calibrated on a sample which had few objects below $10^{9.5-10}\ M_\odot$ -- hence few low $\sigma$ objects anyway).

Recall that $\sigma_a$, and hence $s_e$, is not available for some objects.  Figure~\ref{noV} shows that they tend to have $M_*\le 10^{10} M_\odot$.  Since the IMF for these objects is expected to be close to Chabrier anyway \cite{mangaIMF}, we simply assume that $M_{*}^{\rm \alpha_{JAM}}=M_*$ for these objects.  (Equation~\ref{Mjam} does allow some objects with low but reliable $s_e$ to have $M_{*}^{\rm \alpha_{JAM}}<M_*$; if the missing $s_e$ objects are like these, then we are slightly overestimating their $M_{*}^{\rm \alpha_{JAM}}$ values.)  In any case, as we will see later, $\phi(M_*)$ is relatively flat at low masses, so changing $M_*$ makes little difference.

\subsection{$M_*^{\alpha_{\rm JAM}{\rm FP}}$}\label{sec:fp}

\begin{figure}
 \centering
 \includegraphics[width=8cm]{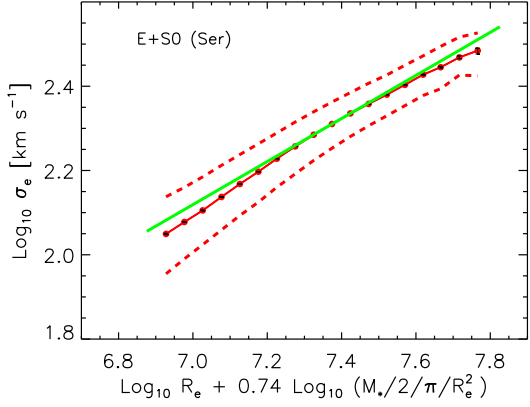}
 \caption{Fundamental Plane proxy for $\sigma_e$ has an rms scatter of 0.075~dex. }
 \label{FP1}
\end{figure}

Our second proxy is motivated by the fact that, especially at high redshifts, $\sigma$ is expensive to estimate directly for large samples.  So it may be necessary to use an observable proxy for $\sigma_e$.  In samples where $R_e$, fixed-IMF $M_*$ and $\sigma_e$ are all available, $\sigma_e$ is seen to be strongly correlated with both $R_e$ and stellar surface brightness $(M_*/2)/(\pi R_e^2)$.  This is expected if galaxies are virialized systems, and is sometimes referred to as a Fundamental Plane correlation \cite{fp87}.

In our SDSS sample of E+S0s, at fixed $R_e$ and stellar surface brightness $(M_*/2)/(\pi R_e^2)$, the velocity dispersion distribution is strongly peaked around a mean value of 
\begin{equation}
 \label{sigmaFP}
  s_{\rm FP}\equiv \langle s_e|r,i\rangle = a r + b i + c,
\end{equation}
where $s\equiv\log_{10}(\sigma/{\rm km~s^{-1}})$, 
$r\equiv\log_{10}(R_e/{\rm kpc})$,
$i\equiv\log_{10}[(M_*/2M_\odot)/\pi (R_e/{\rm kpc})^2]$,
and
 $$(a,b,c) = (0.511\pm 0.010, 0.380\pm 0.008, -1.467\pm 0.073).$$
These coefficients are those values for which the scatter around this mean relation is minimized.  (We used the method described in Sheth \& Bernardi 2012 to account for selection effects and measurement errors.)  Figure~\ref{FP1} shows this projection of the Fundamental Plane.  The `thickness' of the plane -- the rms scatter around equation~(\ref{sigmaFP}) -- is 0.075~dex.  This is the precision with which $R_e$ and fixed-IMF $M_*$ predict $\sigma_{\rm e}$.  

\begin{figure}
 \centering
 \includegraphics[width=8cm]{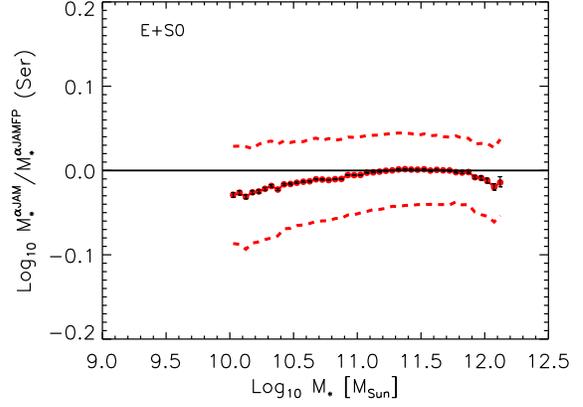}
 \caption{Ratio $M_{*}^{\rm \alpha_{JAM}}/M_{*}^{\rm \alpha_{JAM}FP}$ as a function of $M_{*}$ (fixed to Chabrier IMF). This ratio is unity, and the scatter around it is small, so using a Fundamental Plane motivated proxy in place of the true $\sigma_{\rm e}$ will not lead to a bias in the estimated $M_{*}^{\rm \alpha_{JAM}}$. }
 \label{FP3}
\end{figure}

The result of replacing $s_e$ in equation~(\ref{Mjam}) with this Fundamental Plane estimate $s_{\rm FP}$ is 
\begin{equation}
 \log_{10}\frac{M_*^{\alpha_{\rm JAM}{\rm FP}}}{M_\odot} \equiv \log_{10}\frac{M_*}{M_\odot} + a + b\, (s_{\rm FP} + \Delta_{\rm FP}) + \Delta_{\rm rms}.
 \label{Mjamfp}
\end{equation}
Whereas equation~(\ref{Mjam}) uses $M_*$ and $\sigma_e$, equation~(\ref{Mjamfp}) uses $M_*$ and $R_e$.  Note that $M_*$ itself uses $L$ (which is output by the same analysis of the photometry which returns $R_e$), and $M_*/L$ (which is returned by fitting to stellar-population models).  Figure~\ref{FP3} shows that $M_{*}^{\rm \alpha_{JAM}}/M_{*}^{\rm \alpha_{JAM}FP}\approx 1$ for the majority of the objects.  Evidently, one can use a Fundamental Plane motivated proxy in place of the true $\sigma_e$.  

\subsection{$M_*^{\rm PS-JAM}$}\label{sec:Md}
Our final proxy is $M_*^{\rm PS-JAM}$ of equation~(\ref{Mfinal}), which is motivated by the fact that $M_{*,\rm IMF}\approx M_*^{\rm JAM}$ \cite{califaIMF} and $M_*^{\rm JAM}\approx M_*^{\rm PS-JAM}$ (Figure~\ref{M*final}), and to which we add a Gaussian random number $\Delta_{\rm PS-JAM}$ with rms 0.05~dex to account for the intrinsic scatter which must contribute to Figure~\ref{redoa3d}.  This proxy uses $R_e$ and $\sigma_a$ (aperture corrected to $\sigma_{e/8}$); in this respect, it uses the third of the three possible pairwise combinations of $M_*$, $\sigma$ and $R_e$. 

Whereas this estimate does not make use of $L$, the overall normalization of the surface brightness profile, it does make explicit use of the shape:  both the S{\'e}rsic index $n$ and $R_e$ matter.  Therefore, as part of our study of this proxy, we also consider the deVaucouleur estimate (Figure~\ref{M*deV}), which sets $n=4$ in equation~(\ref{Mdyn}) before inserting in equation~(\ref{Mfinal}).

\section{Mass functions}\label{phiMs}
In effect, Figures~\ref{M*final} and~\ref{M*deV} demonstrate that our (truncated) $M_{*}^{\rm PS}$ estimates are similar to $M_{*}^{\rm \alpha_{JAM}}$, at least for E+S0s.  Therefore, we are finally ready to consider the implications for population statistics.  In all cases, we determine comoving abundances by weighting each object by the comoving volume out to which it could have been observed.  In practice, we account for systematic uncertainties in these $M_*$ estimates by showing a band of allowed values as we describe below.  We do not correct for the broadening due to measurement errors as these are small (Bernardi et al. 2013, 2017a).

\subsection{Truncated S{\'e}rsic and de~Vaucouleurs photometry:  E+S0s}
As we did in the previous section, we first limit the comparison to objects for which equation~(\ref{Mdyn}) is most likely to be accurate: E+S0s.  We select the subsample of objects for which $n \ge 2$ and $p$(E+S0) $\ge 0.7$, and we then weight each object by $p$(E+S0)/$V_{\rm max}$ when computing the mass function.  

\begin{figure}
 \centering
 \includegraphics[width=8cm]{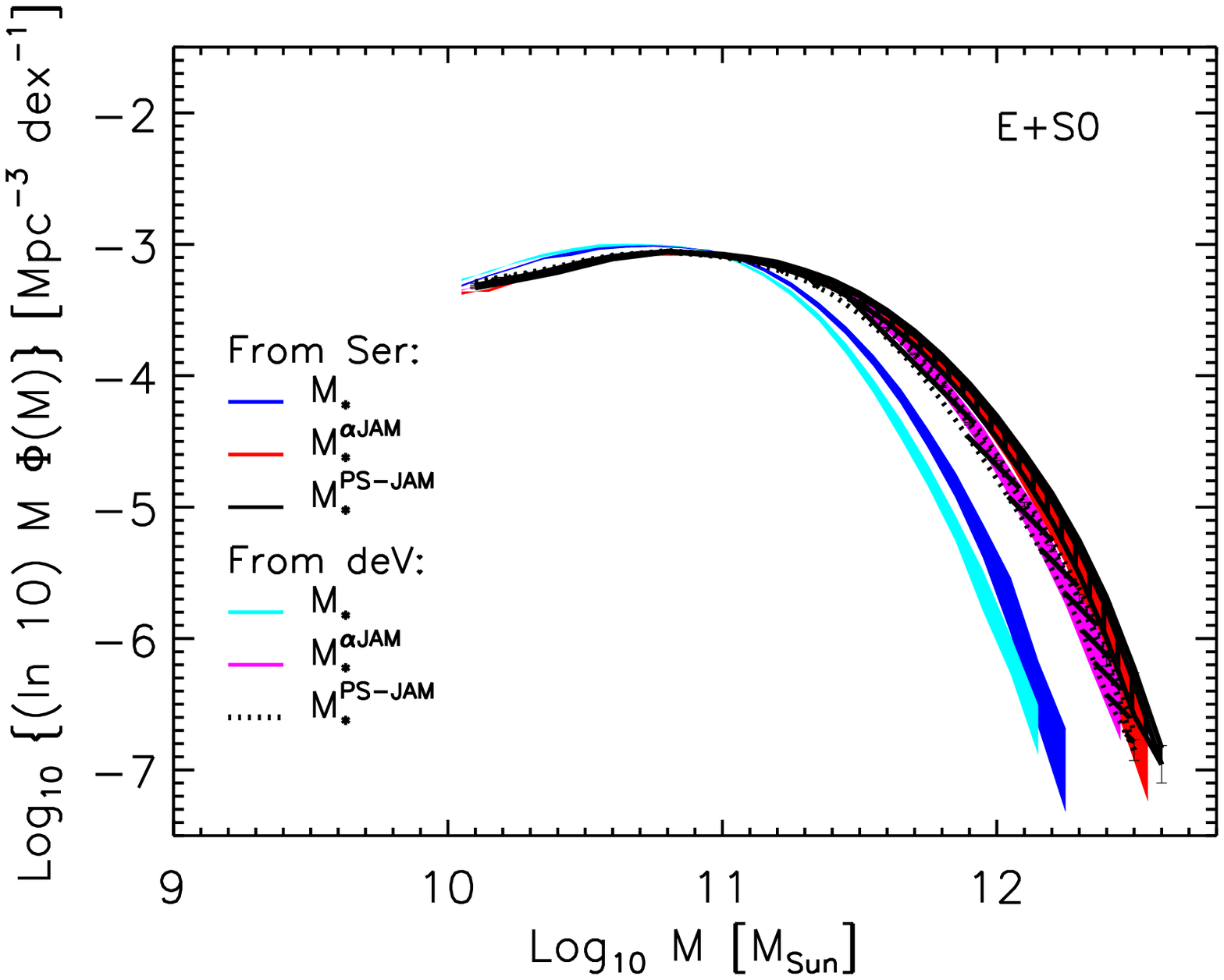}
 \includegraphics[width=8cm]{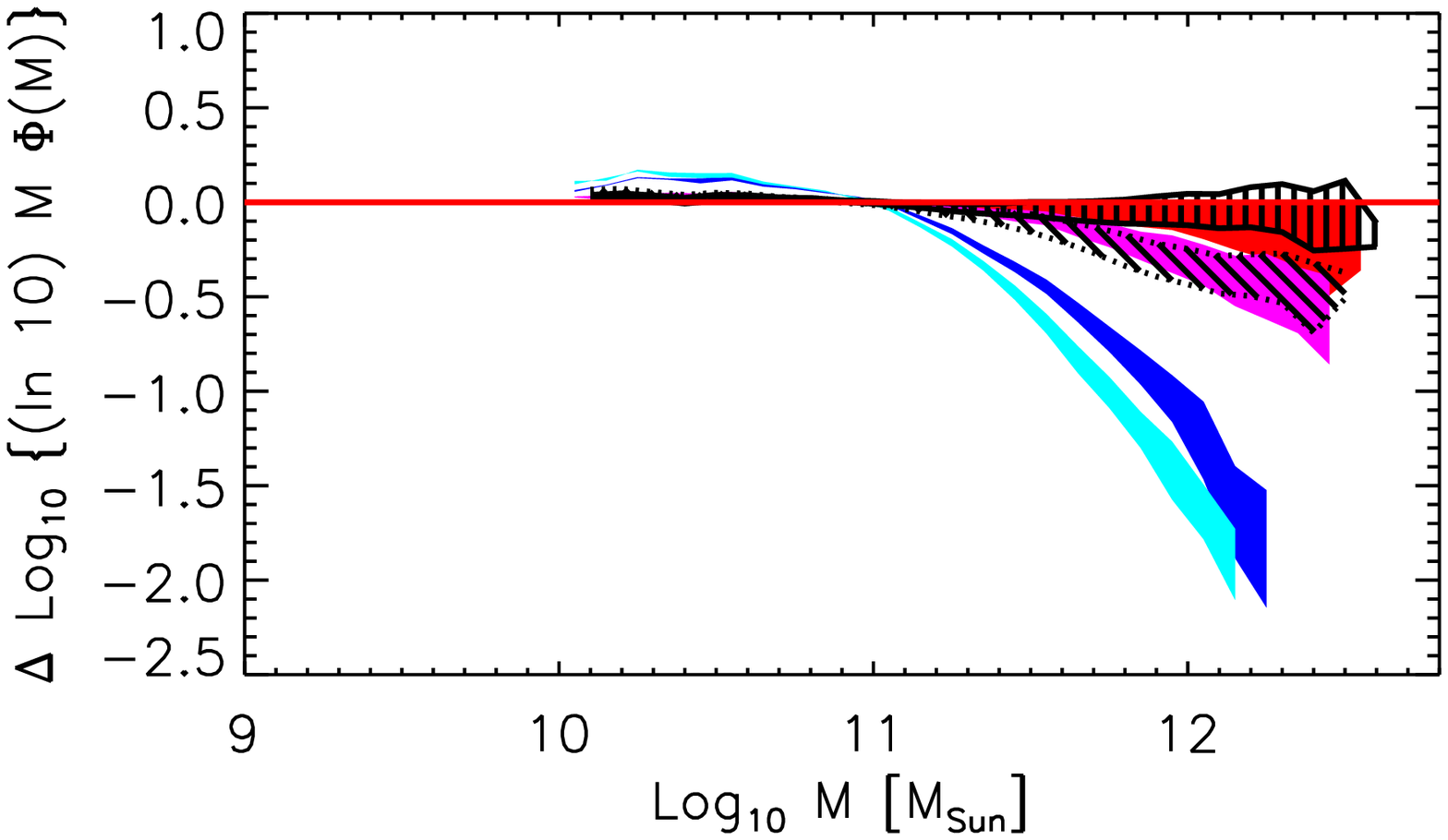}
 \caption{Comparison of $\phi(M_*)$ (Chabrier IMF), $\phi(M_{*}^{\rm \alpha_{JAM}})$ and $\phi(M_{*}^{\rm PS-JAM})$ for E+S0s.  Blue, red and black solid regions are based on single S{\'e}rsic photometry (truncated luminosities); cyan, magenta and black dotted curves on de Vaucouleurs (truncated luminosities).  Top panel shows the mass functions themselves, and bottom panel shows the ratio of the other curves to the top of the red solid region. Blue (cyan) shaded region shows the range bracketed by the dusty and dust-free models of Mendel et al. (2014). Red (magenta) region uses $M_{*}^{\rm \alpha_{JAM}}$ values (equation~\ref{Mjam}) obtained from the top three choices of $(a,b)$ in Table~\ref{tab:IMF}; this range of values transforms the blue (cyan) shaded region to the red (magenta). The black shaded regions show the systematic uncertainty associated with the calibrated $M_*^{\rm PS-JAM}$ values (equation~\ref{Fjam8}).
 }
 \label{F4eso}
\end{figure}

\begin{figure}
 \centering
 \includegraphics[width=8cm]{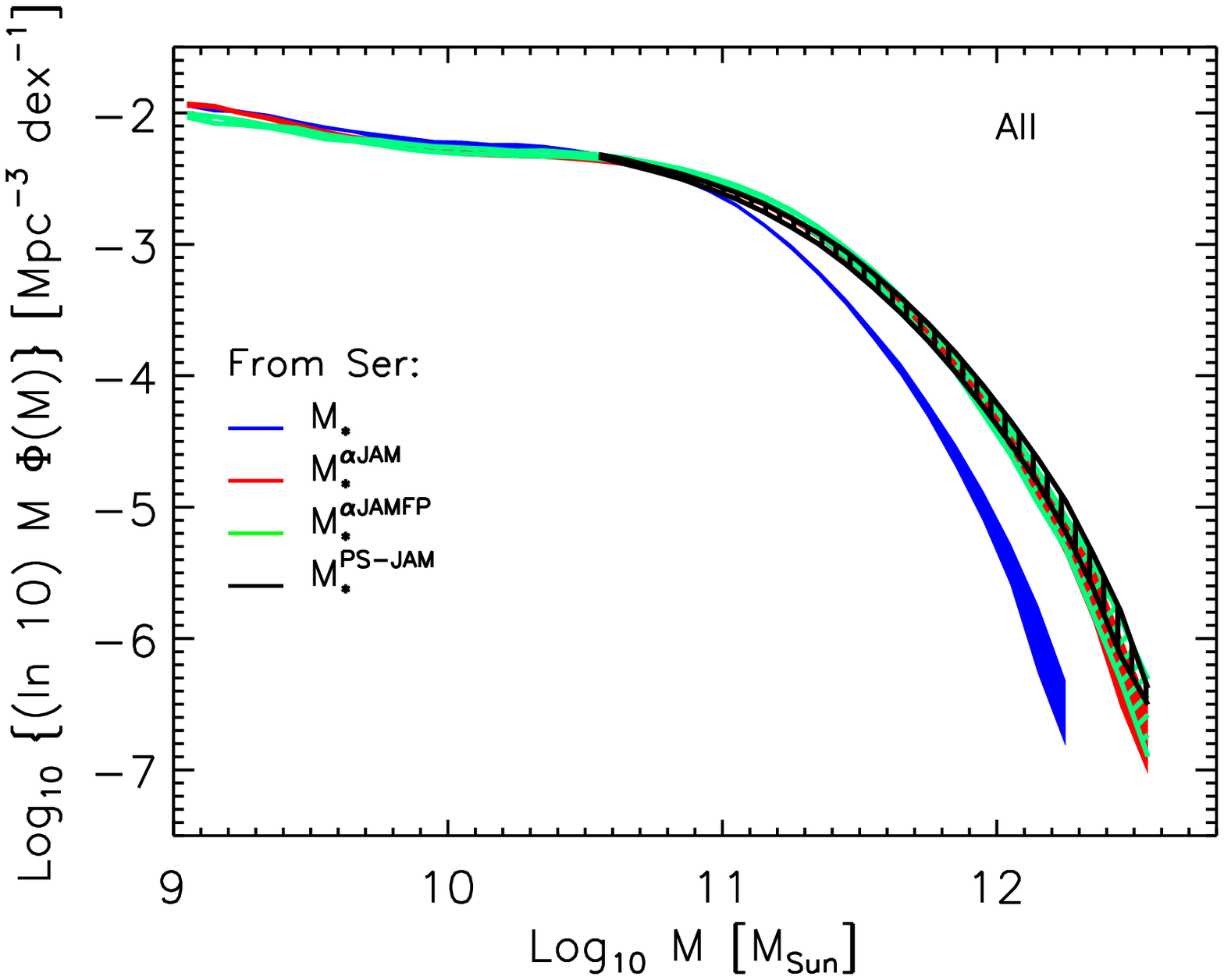}
 \includegraphics[width=8cm]{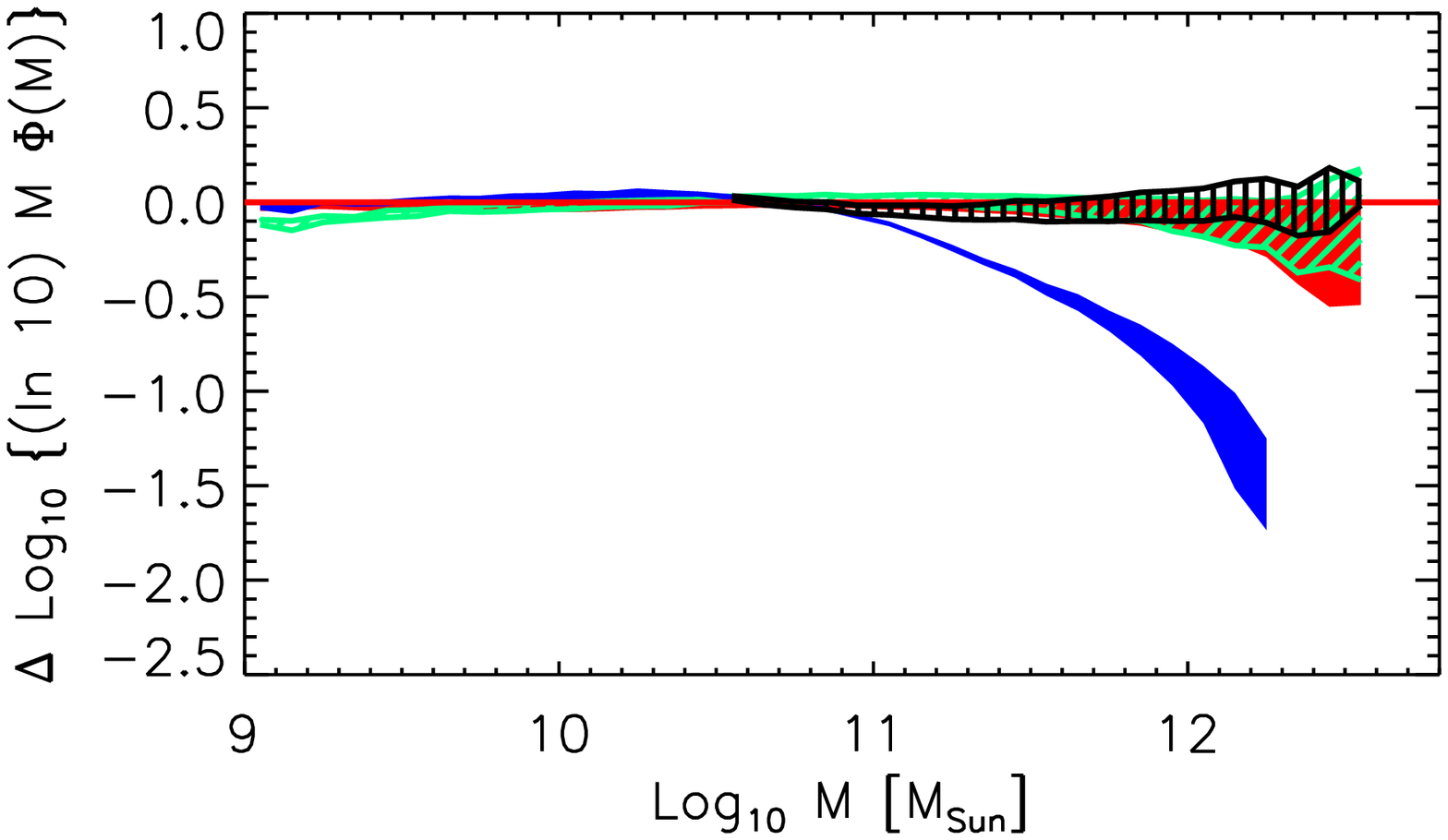} 
 \caption{Comparison of $\phi(M_*)$ (Chabrier IMF), $\phi(M_{*}^{\rm \alpha_{JAM}})$ and $\phi(M_{*}^{\rm \alpha_{JAMFP}})$ (blue, red and green shaded regions) of all galaxies from single S{\'e}rsic photometry (truncated luminosities). Black solid line shows $\phi(M_{*}^{\rm PS-JAM})$ limited to $10^{10.5}M_\odot$, since the measured $\sigma$ is not the appropriate quantity to insert in equation~(\ref{Mdyn}) for the rotationally supported objects which begin to dominate the population at low masses. Table~\ref{tabSer} provides $\phi(M_*)$ and $\phi(M_{*}^{\rm \alpha_{JAM}})$ in tabular form.}
 \label{F4all}
\end{figure}

Figure~\ref{F4eso} compares $\phi(M_*)$ (blue), $\phi(M_{*}^{\rm \alpha_{JAM}})$ (red) and $\phi(M_{*}^{\rm PS-JAM})$ (black) for E+S0s.  The $M_*$ values were obtained by combining truncated S{\'e}rsic luminosity with a range of estimates of $M_*/L$.  The blue shaded region shows the range bracketed by the Mendel et al. dusty and dust-free models (shown in Figure~\ref{F0} below).  The $M_{*}^{\rm \alpha_{JAM}}$ values were obtained from $M_*$ using the top three choices in Table~\ref{tab:IMF} for $(a,b,\Delta_{\rm rms})$ in equation~(\ref{Mjam}); this range of values transforms the blue shaded region to the red.  (Including the scatter $\Delta_{\rm rms}$ reduces the dependence on the exact choice of $(a,b)$.) 

The black shaded region shows $\phi(M_{*}^{\rm PS-JAM})$. The width of this region is given by propagating the uncertainty in $C_R(n)$ (equation~\ref{Fjam8}) as a systematic when we insert it in equation~(\ref{Mfinal}).  
The agreement between the black and red regions is rather good down to masses of order $10^{10}M_\odot$, and should not be surprising given the agreement between $M_{*}^{\rm PS-JAM}$ and $M_{*}^{\rm \alpha_{JAM}}$ shown in Figure~\ref{M*final}.  

For comparison, the cyan, magenta and dotted black regions show the corresponding results based on de~Vaucouleurs photometry.  Since de~Vaucouleurs fits yield systematically lower $L$ (Meert et al. 2015; Bernardi et al. 2017b), both $M_*$ and $M_{*}^{\rm \alpha_{JAM}}$ are smaller, so the cyan and magenta curves lie below the corresponding blue and red ones.  But the important point is that this difference is much smaller than that between the cyan and blue, or the magenta and red, which are shifted with respect to one another by approximately 0.25~dex in $\log_{10}(M_*/M_\odot)$.  Moreover, the black dotted region, which shows the corresponding $M_{*}^{\rm PS-JAM}$ estimate, is in good agreement with the magenta, as expected from the agreement shown in Figure~\ref{M*deV}.

\subsection{Truncated S{\'e}rsic photometry:  All types}
Figure~\ref{F4all} shows that $\phi(M_*^{\alpha_{\rm JAM}})$ (red) and $\phi(M_*^{\rm PS-JAM})$ (black) are in good agreement, and lie systematically above $\phi(M_*)$ (blue), even when we extend our analysis to all galaxy types.  Here, we only show results for truncated S{\'e}rsic photometry, since non-E+S0s are not well-fit by a deVaucouleurs profiles.  In addition, Li et al. (2017) show that the $M_*^{{\rm JAM}}/M_*-\sigma$ relation which we use to estimate $M_*^{\alpha_{\rm JAM}}$ depends slightly on morphological type; although we include this dependence in our estimates, ignoring it makes negligible difference to our results.  This is, in part, because spirals contribute mainly at lower masses where $\phi(M_*)$ is flat, so the correction makes little difference anyway.

We have added a green hashed region, which shows the result of using $M_{*}^{\rm \alpha_{JAM}FP}$ in place of $M_{*}^{\rm \alpha_{JAM}}$.  The agreement between it and the red region shows that the scatter in Figure~\ref{FP3} is small enough that it does not bias the number counts significantly.  Therefore, accounting for IMF-variations when estimating the total stellar mass budget in future surveys of more distant objects may be cheaper than it might otherwise have been.  

We only show $M_{*}^{\rm PS-JAM}$ down to $10^{10.5}M_\odot$, as below this value neither $R_e$ nor $\sigma_e$ (nor the Fundamental Plane proxy for $\sigma_e$) are the appropriate quantities to insert in equation~(\ref{Mfinal}). But above this mass, the agreement with $\phi(M_{*}^{\rm \alpha_{JAM}})$, $\phi(M_{*}^{\rm \alpha_{JAM}FP})$ and $\phi(M_*^{\rm PS-JAM})$ is rather good.  Table~\ref{tabSer} provides these S{\'e}rsic-based $\phi(M_*)$ and $\phi(M_{*}^{\rm \alpha_{JAM}})$ in tabular form. 

\begin{figure}
 \centering
 \includegraphics[width=8cm]{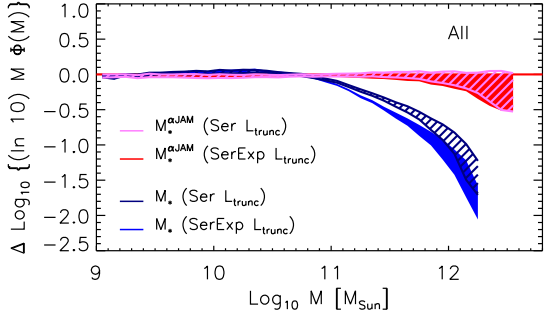}
 \caption{Comparison of $\phi(M_{*})$ (Chabrier IMF) and $\phi(M_{*}^{\rm \alpha_{JAM}})$ computed with truncated S{\'e}rsic and SerExp luminosities.  S{\'e}rsic photometry is slightly brighter, and this leads to slightly larger $\phi(M_*)$.}
 \label{SES}
\end{figure}

\begin{figure}
 \centering
 \includegraphics[width=8cm]{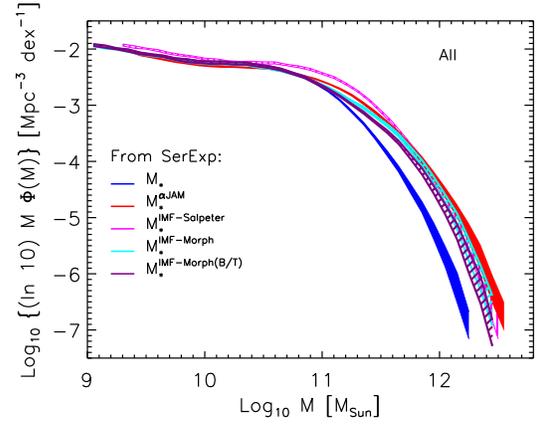}
 \includegraphics[width=8cm]{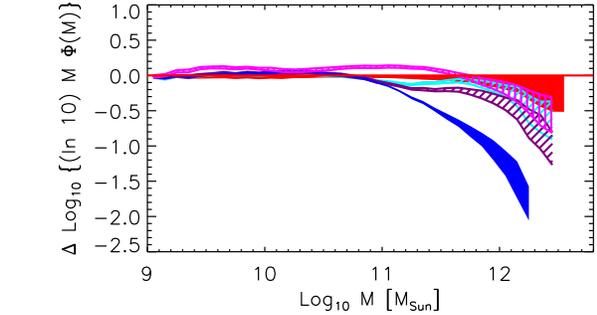}
 \caption{Comparison of $\phi(M_{*}^{\rm \alpha_{JAM}})$ with a variety of stellar population based estimates, now from SerExp photometry (truncated luminosities).  Blue curve uses a Chabrier IMF for all objects, as in the previous Figure; magenta uses a Salpeter IMF instead; cyan uses Salpeter for E+S0s but Chabrier for the rest; and purple uses Salpeter for the bulge component of each galaxy but Chabrier for the other component.}
 \label{F6}
\end{figure}

\subsection{Truncated SerExp photometry: Morphology- or component-dependent IMF}
The previous subsection showed that stellar dynamics-based estimates of the stellar mass function are in good agreement, and predict substantially more mass than a stellar population-based estimate when the IMF is assumed to be Chabrier for all objects.  In this subsection, we will explore a number of other stellar population-based estimates.  For one of them, we would like to allow for the possibility that the bulge and disk components have different IMFs.  Therefore, in this section we use SerExp rather than S{\'e}rsic photometry.  This analysis is useful anyway, since Meert et al. (2013, 2015) and Bernardi et al. (2014) argue that SerExp photometry is the most reliable of the {\tt PyMorph} outputs.  

Figure~\ref{SES}, which is similar in format to the bottom panels of the previous two figures, shows that our mass function determinations based on truncated SerExp parameters are systematically smaller than those based on truncated S{\'e}rsic photometry at the high mass end: the differences are smaller than 0.15~dex except at ${\rm log}_{10} (M_*/M_\odot) > 11.8$ where the differences are $\sim 0.2$~dex.  This is consistent with previous work on truncated S{\'e}rsic photometry (Fischer et al. 2017).  Table~\ref{tabSerExp} provides our SerExp-based $\phi(M_*)$ and $\phi(M_{*}^{\rm \alpha_{JAM}})$ in tabular form. 

Figure~\ref{F6} compares the SerExp-based $\phi(M_{*}^{\rm \alpha_{JAM}})$ (red region) with a variety of stellar population based estimates.  The blue region uses a Chabrier IMF for all objects, as in the previous Figures; it lies below all the others at large masses.  The magenta region uses a Salpeter IMF instead; this boils down to simply shifting the blue curve horizontally by 0.25~dex.  While it provides a good description at high masses, it results in a slight but statistically significant overestimate at all but the largest masses.  This is consistent with the fact that the IMF is closer to Chabrier at low masses.

The cyan region is a crude attempt to allow for IMF variations across the population:  it uses a Salpeter IMF for E+S0s but Chabrier for the rest.  At the highest masses it lies slightly below the red region.  Presumably, the mismatch is because this estimate does not include the fact that the IMF correlates with $\sigma$ even at fixed morphology \cite{mangaIMF}.  

The purple region shows a simple attempt to account for IMF gradients within each galaxy:  it uses Salpeter for the bulge component but Chabrier for the other.  It is rather similar to the cyan region, but still lies noticably below the red.  Of course, in this case the comparison is not quite fair, since the appropriate dynamical-mass estimate should really include the fact that the mass-to-light ratio is not constant in this case.  But we leave this additional complication for future work.

\section{Discussion}
We compared a number of estimates of the dynamical stellar mass in SDSS DR7 galaxies, which were based on S{\'e}rsic fits to the light profile.  One, $M_{*}^{\rm \alpha_{JAM}}$, was based on applying a velocity dispersion dependent correction factor to the fixed-IMF stellar mass estimate $M_*$ (equation~\ref{Mjam}).  This correction was calibrated from a number of previous studies in the literature (Table~\ref{tab:IMF}).  Our second estimate is closely related:  it simply replaces the velocity dispersion in equation~(\ref{Mjam}) with a Fundamental Plane derived proxy (equation~\ref{sigmaFP}).  We showed that doing so does not lead to significant biases (Figure~\ref{FP3}).

We also studied a rather different estimate which was based on estimating a dynamical mass using the measured velocity dispersion in the SDSS fiber, the fitted half-light radius and S{\'e}rsic index $n$.  In its simplest form (equation~\ref{Mdyn}), this estimate assumes that the mass-to-light ratio is constant (i.e., the stellar mass-to-light ratio is constant and dark matter, if present, is assumed to have the same spatial distribution as the stars) and the velocity dispersion isotropic.  These simplifications lead to small biases (Figures~\ref{M*a} and~\ref{M*e8}) which we correct by calibrating to the ATLAS$^{\rm 3D}$ sample (equations~\ref{Mfinal} and \ref{kappa}). Our $\kappa(n, R)$ values (equation~\ref{kappa}) are a modified version of $k(n,R)$ provided by Prugniel \& Simien (1997) (equation~\ref{Mdyn} and Table~\ref{tab:kn}).  As part of this process, we considered the possibility that the shape of the velocity dispersion profile correlates with S{\'e}rsic index (i.e. $(\sigma_R/\sigma_e) = (R/R_e)^{-\gamma(n)}$; equation~\ref{gamman} and Figure~\ref{sigma3d}).  

Our final calibrated dynamical mass estimates, $M_*^{\rm PS-JAM}$, are rather similar to $M_{*}^{\rm \alpha_{JAM}}$ (Figure~\ref{M*final}).  Differences between $M_*^{\rm PS-JAM}$ and $M_{*}^{\rm \alpha_{JAM}}$ are like those which arise from using different relations from Table~\ref{tab:IMF} when estimating $M_{*}^{\rm \alpha_{JAM}}$ in the first place.  The associated stellar mass functions are also similar (Figures~\ref{F4eso} and~\ref{F4all}).  

When $M_*$ is estimated assuming a fixed IMF, then the ratios $M_{*}^{\rm PS-JAM}/M_*$ and $M_{*}^{\alpha_{\rm JAM}}/M_*$ correlate strongly with $\sigma_e$.  Recent work ascribes the $M_{*}^{\alpha_{\rm JAM}}/M_* - \sigma_e$ correlation to the fact that objects with large $\sigma$ tend to have bottom-heavy, dwarf-rich IMFs.
If this is correct, then our $\phi(M_{*}^{\rm PS-JAM})$ or $\phi(M_{*}^{\rm \alpha_{JAM}})$ estimates represent determinations of the stellar mass function which account for the fact that objects with large $\sigma$ tend to have bottom-heavy IMFs.  Doing so dramatically increases the abundance of objects having stellar masses in excess of $10^{11.2} M_\odot$ compared to when the IMF is fixed to Chabrier for all objects (Figures~\ref{F4eso} and~\ref{F4all}).  This increase is qualitatively similar to that which follows from assuming that E+S0s have a Salpeter IMF, whereas other galaxy types are more Chabrier-like (Figure~\ref{F6}).  However, the IMF variation appears to be more closely tied to velocity dispersion than morphology \cite{cvdIMF,califaIMF,mangaIMF}.  This suggests a close connection to the potential well in which a galaxy's stars formed, so our $\sigma$- rather than morphology-based methodology may be closer to the physics which determines the total stellar masses.

Estimating the IMF-dependent stellar mass from detailed spectroscopic features, or from spatially resolved spectroscopy, is prohibitively expensive for a large sample of galaxies such as ours.  However, large samples are necessary to accurately probe the highest masses.  Thus, our methodology has allowed an estimate of `IMF-corrected' stellar mass functions at a fraction of the cost.  In particular, our analysis suggests that future variable-IMF estimates of the total stellar mass could be made as follows.  First, obtain good enough spectra to measure the IMF features for a small subset of the total sample.  Typically, these spectra will also allow a measurement of $\sigma$, from which a correction factor like equation~(\ref{Mjam}) can be calibrated.  Assuming that the photometry is good enough to estimate a S{\'e}rsic-based $n,R_e$ and $L$, one could use this subset to also calibrate a correction factor like equation~(\ref{Mjamfp}) using a Fundamental Plane proxy for $\sigma$.  This calibration could then be applied to all the other objects for which high signal-to-noise spectra are not available.  Since detailed spectra are only required for a small subset of the objects, this vastly reduces the cost of accounting for IMF-variations.  

We end with a note of caution.  When Bernardi et al. (2010) first argued for an increase in $\phi(M_*)$ at high masses, they made the point that this increases the mass scale and weakens the role that must have been played by feedback in regulating star formation.  That was before accounting for IMF-variations.  If the high mass end of $\phi(M_*)$ must be increased further because of IMF effects, then the need for feedback will be shifted to even higher masses.  It is not obvious that this is reasonable.  The increase in $M_*$ is driven by the fact that stellar dynamical masses $M_*^{\rm dyn}$ tend to be larger than those from stellar population modeling (with IMF fixed to Chabrier).  However, these $M_*^{\rm dyn}$ estimates all ignore gradients in the stellar mass-to-light ratio.  We are currently studying if $M_*$ and $M_*^{\rm dyn}$ can be reconciled {\em not} by increasing $M_*$ (as was done here) but by decreasing $M_{\rm dyn}$ (because of gradients).  If gradients can be ignored, then our measurements serve as a benchmark for the $z\approx 0$ census of stellar mass.  We provide them in tabular form for  S{\'e}rsic (Table~\ref{tabSer}) and SerExp (Table~\ref{tabSerExp}) photometric reductions, and hope that this will facilitate comparison of our results with galaxy formation models and simulations.

\subsection*{Acknowledgements}
We are grateful to J. Schaye and B. Clauwens for bringing their work to our attention, to C. Tortora, the referee, for a detailed report, and to the LYTE center for its hospitality when this work was completed.

\appendix
\section{Details associated with $M_*$ and $\sigma$ in the SDSS sample of the main text}\label{ml}

\begin{figure}
 \centering
 \includegraphics[width=8cm]{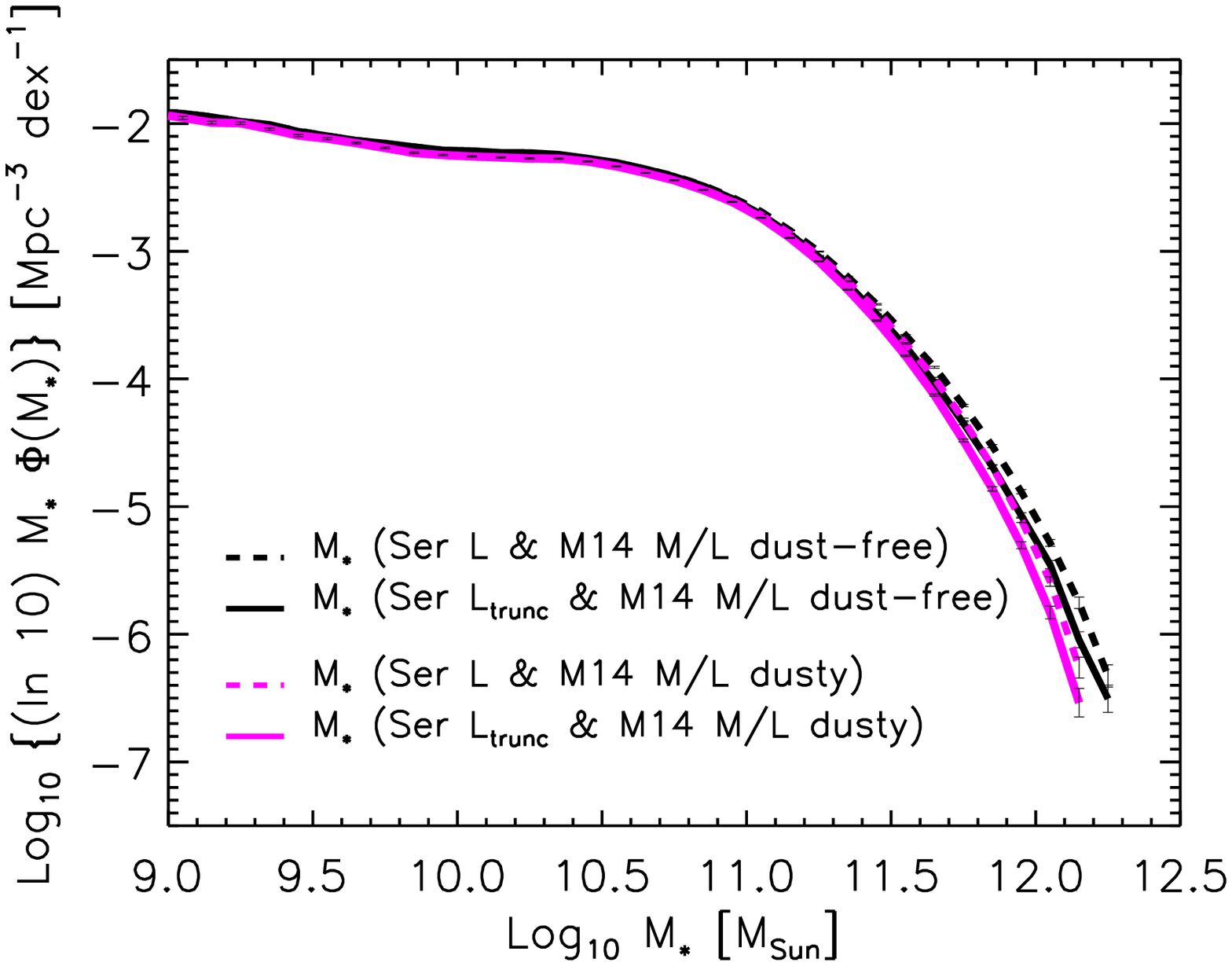}
  \includegraphics[width=8cm]{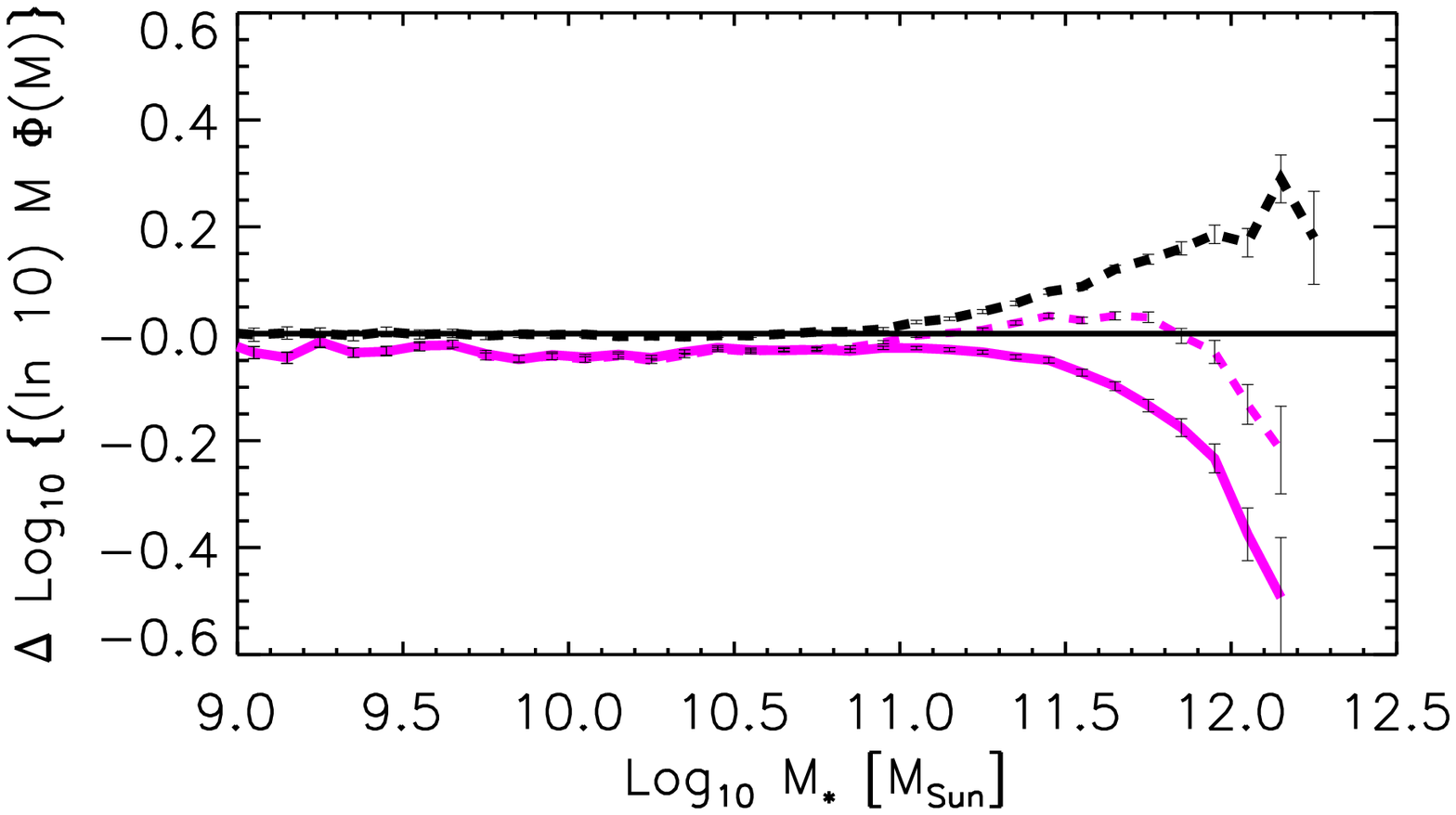}
 \caption{Systematic effects on the stellar mass function.  The luminosity $L$ is from the S{\'e}rsic fit of Meert et al. (2015) and the $M_*/L$ estimates are from Mendel et al. (2014). The black and magenta curves use $M_*/L$ estimates from dust-free and dusty models, respectively. Solid and dashed curves show the effect of truncating the luminosity estimate (solid) or not (dashed). Top panel shows the mass functions themselves, and bottom panel shows the ratio with respect to the dust-free $M_*/L$ and truncated S{\'e}rsic $L$ (black solid line).  We always use truncated $L$ in the main text.}
 \label{F0}
\end{figure}

\begin{figure}
 \centering
 \includegraphics[width=8cm]{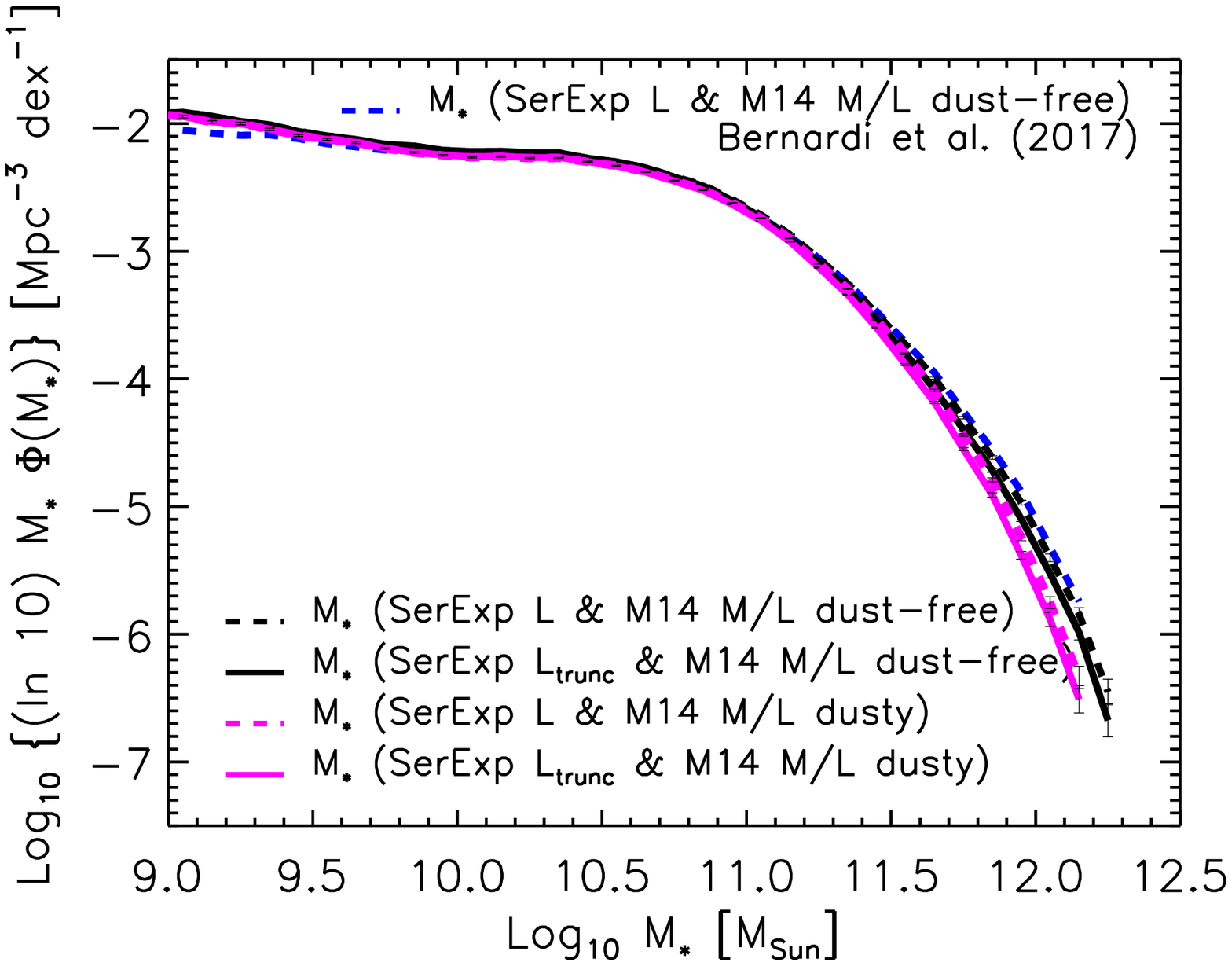}
  \includegraphics[width=8cm]{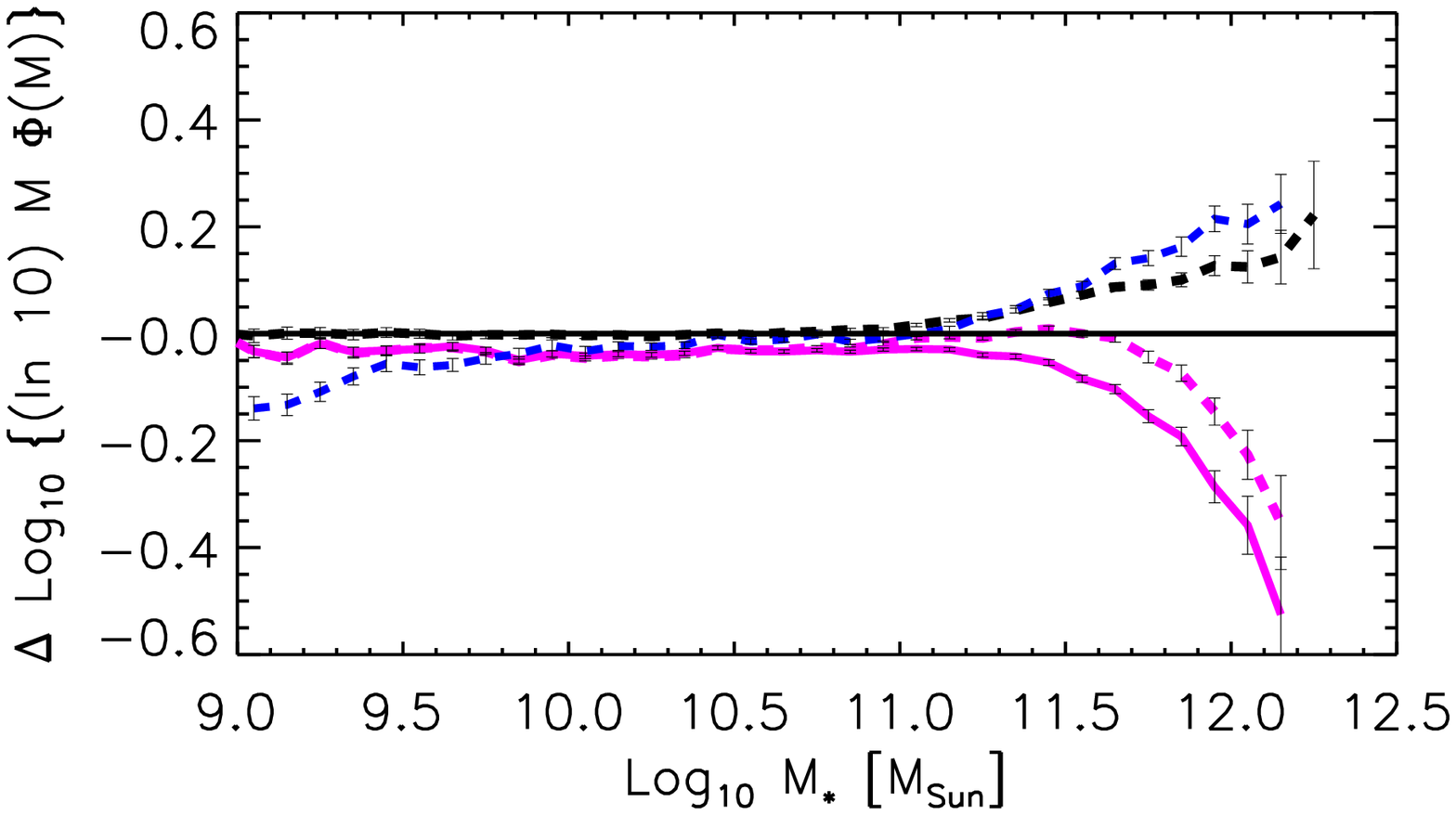}
 \caption{Systematic effects on the stellar mass function.  The luminosity $L$ is from the SerExp fit of Meert et al. (2015) and the $M_*/L$ estimates are from Mendel et al. (2014). The two lowest curves use $M_*/L$ estimates from dusty models; they show the effect of truncating the luminosity estimate (solid magenta) or not (dashed magenta).  Dashed black curve uses $M_*/L$ from a dust-free model with truncated luminosity (solid) or not (dashed), and dashed blue curve shows the estimate from Bernardi et al. (2017a).  Top panel shows the mass functions themselves, and bottom panel shows the ratio with respect to the dust-free $M_*/L$ and SerExp $L$ (black dashed line).}
 \label{F1}
\end{figure}

\begin{figure}
 \centering
 \includegraphics[width=8cm]{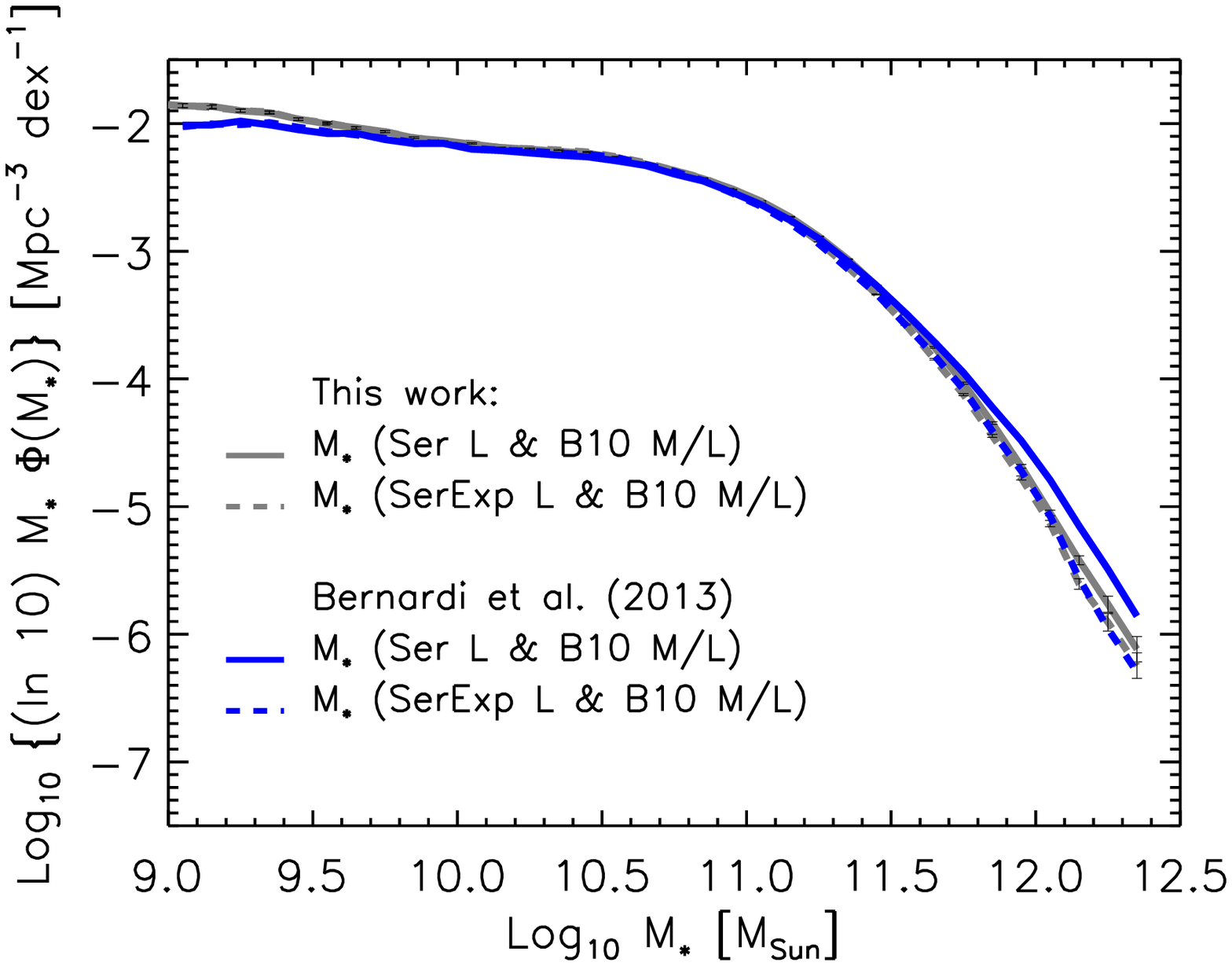}
 \includegraphics[width=8cm]{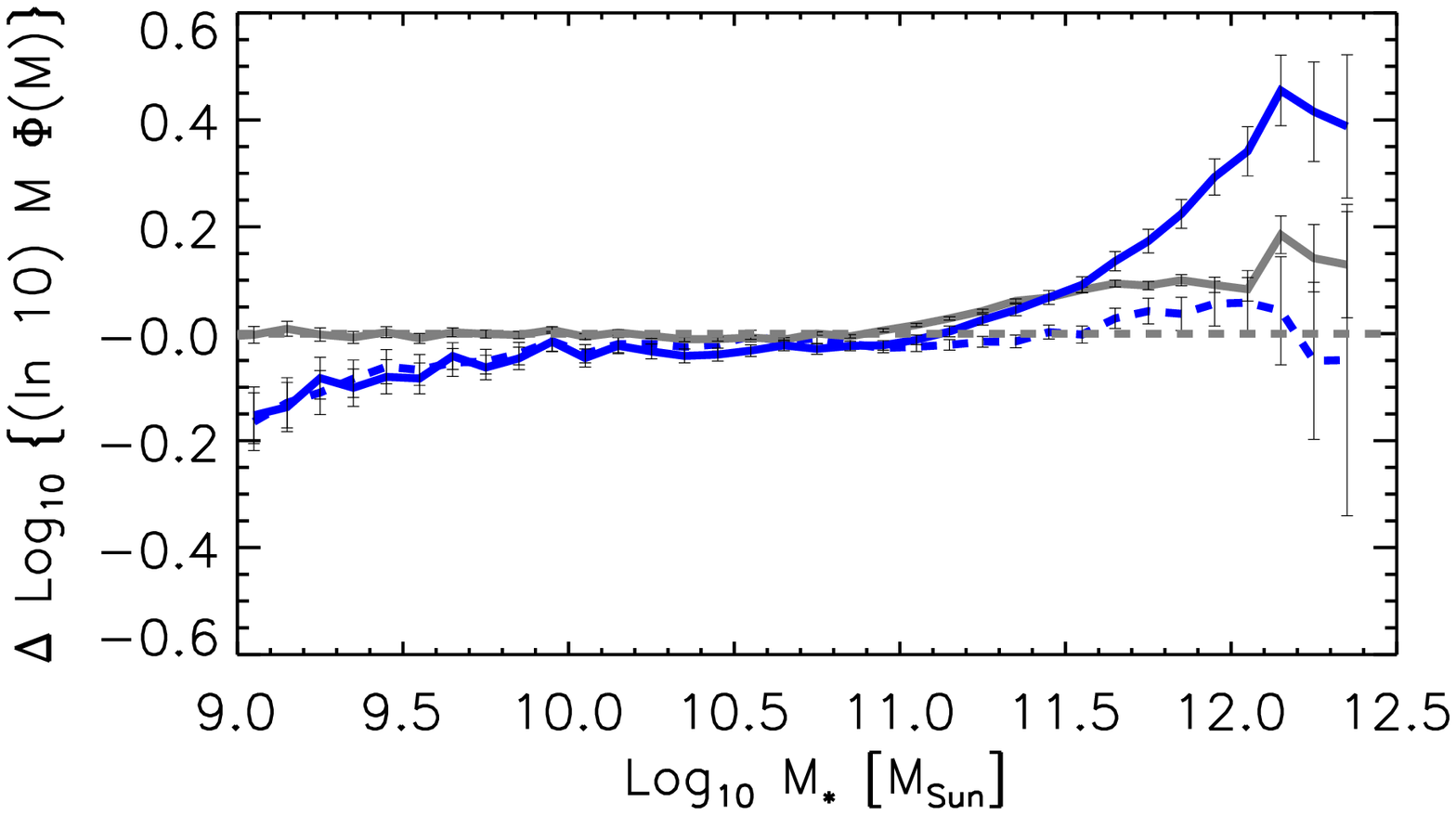}
 \caption{Systematic differences with respect to Bernardi et al. (2013).  The SerExp differences (compare dashed curves) are because the full DR7 sample we use in this work is larger than the subsample used in Bernardi et al. 2013 (which was the same as that in Bernardi et al. 2010).  Sampling errors give rise to the differences at the lowest masses. The S{\'e}rsic differences have an additional effect:  whereas we eliminate objects which Meert et al. (2015) flag as being bad fits, Bernardi et al. (2013) used a preliminary version of the flags.  Removing the objects with published (rather than preliminary) flags reduces the difference between the S{\'e}rsic- and SerExp-based mass functions.}
 \label{compareB13}
\end{figure}

\subsection{Systematic effects when the IMF is fixed}
One of the first steps in the main text is to use stellar mass estimates $M_*$ which assume that the IMF is constant across the population.  Bernardi et al. (2017a) provide an extensive discussion of the systematic uncertainties associated with these estimates.  These have two sources: those from $L$ and others from $M_*/L$. Bernardi et al. (2017a) make the case that there is now general agreement that the $L$ estimates based on single component S{\'e}rsic or two component SerExp photometry are reliable. Compared to these more recent analyses, previous work based on the SDSS pipeline photometry leads to underestimates of $\rho_*(\ge\ M_*)$ by factors of $3-10$ in the mass range $10^{11} - 10^{11.6}M_\odot$, but up to a factor of 100 at higher stellar masses. Bernardi et al. (2017a) show that systematics in photometry now amount to only about 0.1~dex in the stellar mass density -- this is a significant improvement with respect to a decade ago.  

There is on-going discussion on whether one should assume the profile extends to infinity when estimating the total $L$, and if not, how the profile should be truncated (for details see Bernardi et al. 2017a and Fischer et al. 2017).  However, this only leads to small differences in $\phi(M_*)$.  The solid and dashed curves in Figures~\ref{F0} and~\ref{F1} show the effect of truncating or not the S{\'e}rsic and SerExp estimates of the total luminosity.  These differences are comparable to or smaller than those arising from different assumptions about the nature of the stellar population -- which give rise to systematic changes in $M_*/L$ -- and matter most at the high mass end.  We illustrate this by contrasting dusty and dust free models from Mendel et al. (2014).  In the main text we always work with truncated luminosities, and we use the range between the dashed magenta and dashed black curves to represent the systematic uncertainty on $M_*/L$ when the IMF is fixed.  Note that this is actually an underestimate of the uncertainty in stellar population model-based $M_*$ values; for the full range of SP-related systematic uncertainties, see Bernardi et al. (2017a).  

For completeness, the dashed blue curve in Figure~\ref{F1} shows the counts reported by Bernardi et al. (2017a; they used the total luminosity, not truncated).  We expect small differences because we use the full DR7 ($\sim 8000$~deg$^2$), whereas they only used a subset ($\sim 4700$~deg$^2$) -- the same subset used in Bernardi et al. (2010; 2013).  The difference is slightly more than the increased area, since galaxies which were missed to fiber collisions in earlier runs have since been observed with SDSS plates, so the survey completeness has increased.  As a result, whereas  Bernardi et al. (2010; 2013; 2017a) multiplied their $V_{\rm max}^{-1}$ estimates of the measured comoving densities by a factor of $0.93^{-1}$, we now only use $0.98^{-1}$ (see discussion in Section 2.6 in Thanjavur et al. 2016).  Comparison with the blue curve shows there is little difference in the SerExp counts.

\begin{figure}
 \centering
\includegraphics[width=8cm]{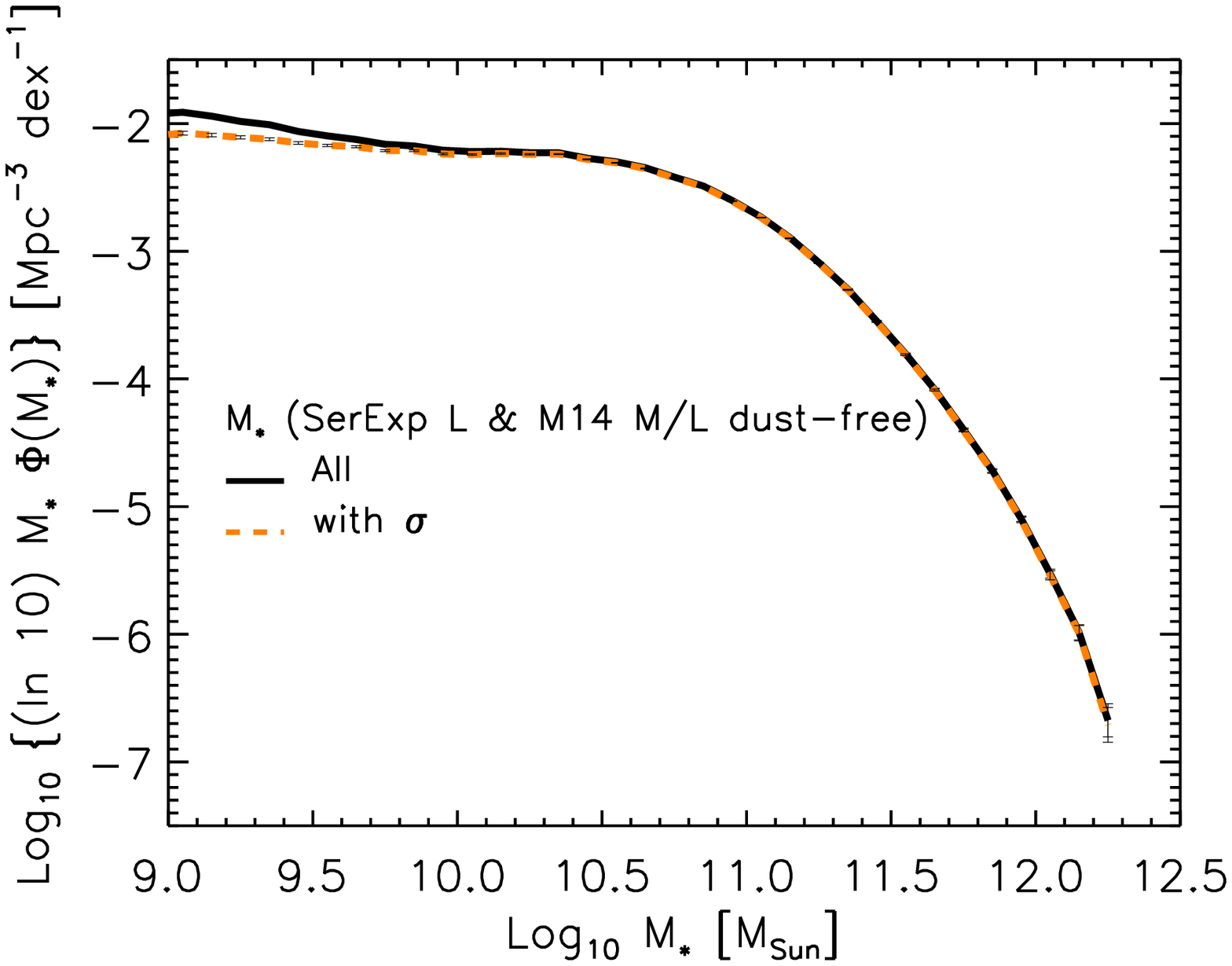}
 \caption{Stellar mass function of the full SDSS Main Galaxy Sample (solid, same as black solid curve in Figure~\ref{F1}) and the subset with measured velocity dispersions (dashed).  Differences only appear below $10^{10}M_\odot$.  }
 \label{noV}
\end{figure}

Finally, we compare with Bernardi et al. (2013).  In this case, in addition to the sample size, there are differences in $M_*/L$ as well as in how bad fits are identified.  We eliminate objects which were flagged as being bad fits in the published version of Meert et al. (2015), whereas Bernardi et al. (2013) eliminated objects based on a preliminary version of the flags.  Figure~\ref{compareB13} shows the SerExp and S{\'e}rsic-based mass functions from Bernardi et al. (2013), based on their smaller sample, and our results from the full DR7 sample used in the main text. Here we use the same $M_*/L$ as Bernardi et al. (2013; i.e. the $M_*/L$ defined in Bernardi et al. 2010) instead of the $M_*/L$ estimates from Mendel et al. (2014). The SerExp counts are similar:  the difference between the dashed lines is small, and similar to the difference between the blue and black dashed curves in Figure~\ref{F1}.  However, the same is not true for S{\'e}rsic-based counts.  Since the SerExp analysis indicates that the $M_*/L$ differences are not large, this difference is driven by how bad fits were flagged.  Using the published flags, as we do in the main text, reduces the S{\'e}rsic counts at the highest masses (compare solid lines), bringing them closer to the SerExp counts.

\subsection{Effect of missing $\sigma$}
The main text uses equation~(\ref{Mjam}) to transform fixed-IMF $M_*$ values to $M_{*}^{\rm \alpha_{JAM}}$ values which are expected to account for variations in the IMF.  This requires an estimate of the velocity dispersion $\sigma_e$, which we determine by applying a small correction to $\sigma_{\rm a}$.  However, $\sigma_{\rm a}$ is not available for a small subset of the DR7 Main Galaxy Sample.  Figure~\ref{noV} shows that these objects are primarily at small $M_*$, so we expect them to have small $\sigma_{\rm a}$, as a result of which we expect $M_{*}^{\rm \alpha_{JAM}}\approx M_*$ for these objects (c.f. Figure~\ref{M*a} in the main text).  Therefore, we simply assume that $M_{*}^{\rm \alpha_{JAM}}=M_*$ for these objects. (As we said in the main text, equation~\ref{Mjam} does allow some objects with low but reliable $\sigma$ to have $M_{*}^{\rm \alpha_{JAM}}<M_*$; if the missing $\sigma$ objects are like these, then we are slightly overestimating their $M_{*}^{\rm \alpha_{JAM}}$ values.  It is worth bearing in mind, however, that equation~\ref{Mjam} was calibrated on a sample in which there were few objects with $M_* < 10^9M_\odot$.) Moreover, at low masses, $\phi(M_*)$ is relatively flat, so changing $M_*$ makes little difference to $\phi(M_*)$. Thus, this figure shows that although some objects do not have $\sigma_{\rm a}$ they are not the massive galaxies which are of most interest in the main text.

\section{Mass functions in tabular form}\label{tables}
This Appendix provides the mass functions shown in Figures~\ref{F4all} and~\ref{F6} in tabular form.  

\begin{table}
 \centering
 \caption{Table of observed stellar mass functions $\Phi\equiv\log_{10}[\ln(10)M_*\phi(M_*)]$, in units of Mpc$^{-3}$dex$^{-1}$, as a function of ${\cal M}_*\equiv \log_{10}(M_*/M_\odot)$ with truncated S{\'e}rsic photometry from Meert et al. (2015) and $M_*/L$ from Mendel et al. (2014; M14). M14$_{\rm d}$ and M14$_{\rm df}$ are the values associated with dusty and dust-free models (the bottom and top boundaries of the blue region in Figure~\ref{F4all}).  $\alpha_{\rm JAM} - {\rm M14}$ values were obtained by transforming these M14 stellar masses using the top three choices in Table~\ref{tab:IMF} for $(a,b)$ in equation~(\ref{Mjam}). Here we report the bottom and top boundaries (red region in Figure~\ref{F4all}).}
 \begin{tabular}{ccccc}
 \hline
  ${\cal M}_*$ & $\Phi^{\rm M14_{\rm d}}_{\rm Ser}$ & $\Phi^{\rm M14_{\rm df}}_{\rm Ser}$ & $\Phi^{\alpha_{\rm JAM}-{\rm M14}_{\rm bot}}_{\rm Ser}$ & $\Phi^{\alpha_{\rm JAM}-{\rm M14}_{\rm top}}_{\rm Ser}$\\
 \hline
$ 9.05$ & $ -1.955$ & $ -1.922$ & $ -1.960$ & $ -1.914$ \\
$ 9.15$ & $ -1.993$ & $ -1.946$ & $ -1.971$ & $ -1.931$ \\
$ 9.25$ & $ -1.999$ & $ -1.982$ & $ -2.021$ & $ -1.985$ \\
$ 9.35$ & $ -2.043$ & $ -2.011$ & $ -2.062$ & $ -2.020$ \\
$ 9.45$ & $ -2.092$ & $ -2.058$ & $ -2.122$ & $ -2.072$ \\
$ 9.55$ & $ -2.121$ & $ -2.098$ & $ -2.163$ & $ -2.124$ \\
$ 9.65$ & $ -2.150$ & $ -2.131$ & $ -2.206$ & $ -2.167$ \\
$ 9.75$ & $ -2.196$ & $ -2.157$ & $ -2.241$ & $ -2.192$ \\
$ 9.85$ & $ -2.229$ & $ -2.183$ & $ -2.284$ & $ -2.229$ \\
$ 9.95$ & $ -2.250$ & $ -2.209$ & $ -2.306$ & $ -2.259$ \\
$10.05$ & $ -2.262$ & $ -2.214$ & $ -2.328$ & $ -2.276$ \\
$10.15$ & $ -2.269$ & $ -2.231$ & $ -2.337$ & $ -2.290$ \\
$10.25$ & $ -2.279$ & $ -2.232$ & $ -2.345$ & $ -2.305$ \\
$10.35$ & $ -2.282$ & $ -2.247$ & $ -2.349$ & $ -2.310$ \\
$10.45$ & $ -2.301$ & $ -2.274$ & $ -2.364$ & $ -2.331$ \\
$10.55$ & $ -2.339$ & $ -2.308$ & $ -2.382$ & $ -2.350$ \\
$10.65$ & $ -2.385$ & $ -2.356$ & $ -2.406$ & $ -2.378$ \\
$10.75$ & $ -2.442$ & $ -2.411$ & $ -2.445$ & $ -2.416$ \\
$10.85$ & $ -2.514$ & $ -2.484$ & $ -2.495$ & $ -2.466$ \\
$10.95$ & $ -2.603$ & $ -2.578$ & $ -2.547$ & $ -2.515$ \\
$11.05$ & $ -2.714$ & $ -2.689$ & $ -2.627$ & $ -2.588$ \\
$11.15$ & $ -2.864$ & $ -2.837$ & $ -2.714$ & $ -2.674$ \\
$11.25$ & $ -3.037$ & $ -3.003$ & $ -2.825$ & $ -2.778$ \\
$11.35$ & $ -3.236$ & $ -3.200$ & $ -2.960$ & $ -2.904$ \\
$11.45$ & $ -3.461$ & $ -3.416$ & $ -3.126$ & $ -3.062$ \\
$11.55$ & $ -3.721$ & $ -3.658$ & $ -3.306$ & $ -3.227$ \\
$11.65$ & $ -3.996$ & $ -3.909$ & $ -3.508$ & $ -3.422$ \\
$11.75$ & $ -4.317$ & $ -4.209$ & $ -3.742$ & $ -3.635$ \\
$11.85$ & $ -4.690$ & $ -4.527$ & $ -4.002$ & $ -3.878$ \\
$11.95$ & $ -5.103$ & $ -4.883$ & $ -4.296$ & $ -4.135$ \\
$12.05$ & $ -5.587$ & $ -5.285$ & $ -4.615$ & $ -4.418$ \\
$12.15$ & $ -6.261$ & $ -5.754$ & $ -4.964$ & $ -4.744$ \\
$12.25$ & $ -6.810$ & $ -6.327$ & $ -5.366$ & $ -5.076$ \\
$12.35$ & $  --  $ & $  --  $ & $ -5.869$ & $ -5.436$ \\
$12.45$ & $  --  $ & $  --  $ & $ -6.523$ & $ -5.970$ \\
$12.55$ & $  --  $ & $  --  $ & $ -7.030$ & $ -6.486$ \\
 \hline
 \end{tabular}
 \label{tabSer}
\end{table}

\begin{table}
 \centering
 \caption{Same as Table~\ref{tabSer}, but for truncated SerExp photometry from Meert et al. (2015) (blue and red regions in Figure~\ref{F6}).}
 \begin{tabular}{ccccc}
   \hline
  ${\cal M}_*$ & $\Phi^{\rm M14_{\rm d}}_{\rm SerExp}$ & $\Phi^{\rm M14_{\rm df}}_{\rm SerExp}$ & $\Phi^{\alpha_{\rm JAM}-{\rm M14}_{\rm bot}}_{\rm SerExp}$ & $\Phi^{\alpha_{\rm JAM}-{\rm M14}_{\rm top}}_{\rm SerExp}$\\   
  \hline
$ 9.05$ & $ -1.944$ & $ -1.911$ & $ -1.948$ & $ -1.914$ \\
$ 9.15$ & $ -1.986$ & $ -1.942$ & $ -1.985$ & $ -1.931$ \\
$ 9.25$ & $ -2.002$ & $ -1.983$ & $ -2.016$ & $ -1.987$ \\
$ 9.35$ & $ -2.043$ & $ -2.009$ & $ -2.055$ & $ -2.009$ \\
$ 9.45$ & $ -2.093$ & $ -2.062$ & $ -2.127$ & $ -2.082$ \\
$ 9.55$ & $ -2.125$ & $ -2.096$ & $ -2.166$ & $ -2.123$ \\
$ 9.65$ & $ -2.148$ & $ -2.125$ & $ -2.207$ & $ -2.166$ \\
$ 9.75$ & $ -2.193$ & $ -2.162$ & $ -2.247$ & $ -2.196$ \\
$ 9.85$ & $ -2.225$ & $ -2.176$ & $ -2.276$ & $ -2.233$ \\
$ 9.95$ & $ -2.247$ & $ -2.209$ & $ -2.306$ & $ -2.252$ \\
$10.05$ & $ -2.261$ & $ -2.219$ & $ -2.324$ & $ -2.278$ \\
$10.15$ & $ -2.255$ & $ -2.217$ & $ -2.336$ & $ -2.285$ \\
$10.25$ & $ -2.267$ & $ -2.228$ & $ -2.337$ & $ -2.298$ \\
$10.35$ & $ -2.265$ & $ -2.228$ & $ -2.348$ & $ -2.310$ \\
$10.45$ & $ -2.298$ & $ -2.272$ & $ -2.357$ & $ -2.326$ \\
$10.55$ & $ -2.332$ & $ -2.299$ & $ -2.371$ & $ -2.340$ \\
$10.65$ & $ -2.379$ & $ -2.346$ & $ -2.408$ & $ -2.374$ \\
$10.75$ & $ -2.451$ & $ -2.420$ & $ -2.444$ & $ -2.418$ \\
$10.85$ & $ -2.523$ & $ -2.489$ & $ -2.496$ & $ -2.462$ \\
$10.95$ & $ -2.632$ & $ -2.602$ & $ -2.557$ & $ -2.519$ \\
$11.05$ & $ -2.760$ & $ -2.732$ & $ -2.633$ & $ -2.597$ \\
$11.15$ & $ -2.923$ & $ -2.894$ & $ -2.729$ & $ -2.685$ \\
$11.25$ & $ -3.128$ & $ -3.088$ & $ -2.844$ & $ -2.792$ \\
$11.35$ & $ -3.340$ & $ -3.298$ & $ -2.994$ & $ -2.933$ \\
$11.45$ & $ -3.601$ & $ -3.547$ & $ -3.154$ & $ -3.088$ \\
$11.55$ & $ -3.888$ & $ -3.804$ & $ -3.341$ & $ -3.263$ \\
$11.65$ & $ -4.185$ & $ -4.081$ & $ -3.550$ & $ -3.457$ \\
$11.75$ & $ -4.550$ & $ -4.395$ & $ -3.777$ & $ -3.679$ \\
$11.85$ & $ -4.909$ & $ -4.717$ & $ -4.047$ & $ -3.903$ \\
$11.95$ & $ -5.382$ & $ -5.096$ & $ -4.338$ & $ -4.174$ \\
$12.05$ & $ -5.884$ & $ -5.526$ & $ -4.665$ & $ -4.461$ \\
$12.15$ & $ -6.510$ & $ -5.985$ & $ -5.005$ & $ -4.766$ \\
$12.25$ & $ -7.159$ & $ -6.674$ & $ -5.449$ & $ -5.101$ \\
$12.35$ & $  --  $ & $  --  $ & $ -5.930$ & $ -5.484$ \\
$12.45$ & $  --  $ & $  --  $ & $ -6.533$ & $ -6.019$ \\
$12.55$ & $  --  $ & $  --  $ & $ -7.020$ & $ -6.504$ \\
  \hline
 \end{tabular}
 \label{tabSerExp}
\end{table}

\end{document}